\documentclass[twocolumn,floatfix, superscriptaddress,pre,aps,a4paper]{revtex4}

\usepackage{graphicx}
\usepackage{amssymb}
\usepackage{amsmath}
\usepackage[utf8]{inputenc}
\usepackage{cancel}
\usepackage[usenames,dvipsnames]{color}
\usepackage{psfrag}
\usepackage{epsfig}
\usepackage{bbm}
\usepackage{bm}
\usepackage{dsfont}
\usepackage{soul}
\usepackage{colonequals}
\usepackage{float}
\usepackage{enumitem}
\usepackage{appendix}
\usepackage{hyperref}

\begin{document}

\title{Nonreciprocal nanoparticle refrigerators: design principles and constraints}
\author{Sarah A.M. Loos}
\email{sl2127@cam.ac.uk}
\address{DAMTP, University of Cambridge, Wilberforce Road, Cambridge CB3 0WA, United Kingdom}
\address{ICTP -- International Centre for Theoretical Physics, Strada Costiera, 11, 34151 Trieste, Italy}
\author{Saeed Arabha}
\address{Department of Mechanical Engineering, Lassonde School of Engineering, York University, Toronto, Canada}
\address{Advanced Simulation and Computing Laboratory (ASCL), Imam Khomeini International University, Qazvin, Iran}
\author{Ali Rajabpour}
\address{Advanced Simulation and Computing Laboratory (ASCL), Imam Khomeini International University, Qazvin, Iran}
\author{Ali Hassanali}
\address{ICTP -- International Centre for Theoretical Physics, Strada Costiera, 11, 34151 Trieste, Italy}
\author{\'Edgar Rold\'an}
\address{ICTP -- International Centre for Theoretical Physics, Strada Costiera, 11, 34151 Trieste, Italy}
\date{\today}

\begin{abstract}
We study the heat transfer between two  nanoparticles held at different temperatures that interact through nonreciprocal forces, by combining molecular dynamics simulations with stochastic thermodynamics. 
Our simulations  reveal that it is possible to construct nano refrigerators that generate a net heat transfer from a cold to a hot reservoir  at the expense of power exerted by the nonreciprocal forces. Applying concepts from stochastic thermodynamics  to a minimal underdamped Langevin model, we derive exact analytical expressions predictions for the fluctuations of work, heat, and efficiency, which reproduce thermodynamic quantities extracted from the molecular dynamics simulations. The theory only involves a single unknown parameter, namely an effective friction coefficient, which we estimate fitting the results of the molecular dynamics simulation to our theoretical predictions. Using this framework, we also establish design principles which identify the minimal amount of entropy production that is needed to achieve a certain amount of uncertainty in the power fluctuations of our nano refrigerator. Taken together, our results shed light on how the direction and fluctuations of heat flows in natural and artificial nano machines can be accurately quantified and controlled by using nonreciprocal forces. 
\end{abstract}

\maketitle

\section{Introduction}
    
Experimental techniques in single-molecule optical trapping and biophysics allow to extract  real-time information  of the state of a nanosystem with exquisite precision~\cite{bustamante2005nonequilibrium,ashkin1986observation,haroche1991trapping}.  Such information is commonly used to infer both thermodynamical and dynamical properties through data-analysis techniques.   
Alongside,  as inspired by Maxwell's demon thought experiment, information acquired from a nanosystem  can be delivered into  work by executing feedback-control protocols~\cite{bechhoefer2005feedback,toyabe2010experimental,campisi2017feedback,ciliberto2020autonomous,parrondo1996criticism}. 
In parallel to experimental progress, the development of  stochastic thermodynamics (ST) over the last two decades provides a robust  theoretical framework to  describe accurately information-to-work transduction that takes into account  nanoscale   fluctuations~\cite{Sekimoto2010,jarzynski2011equalities,Seifert2012,van2015ensemble,peliti2021stochastic}. Combining  stochastic thermodynamics and feedback-cooling techniques has attracted attention towards refrigerating capabilities of small systems under nonequilibrium conditions~\cite{gieseler2012subkelvin,gieseler2014dynamic}.

An important step to optimize the design of microscopic refrigerators  is to bridge the gap between theoretical proposals and experiments 
through the powerful method of all-atoms simulations. Nonequilibrium Molecular Dynamics (MD) studies  provide a suitable platform for the study of heat transfer and fluctuations at the nanoscale~\cite{rajabpour2019thermal,todd2017nonequilibrium, roodbari2022interfacial, muller1997simple, allen2017computer}.  However, little is known yet about the design and performance of information demons at the atomic scale. In particular, are there generic principles that constraint  the forces needed to ensure a prescribed value for the heat transfer between two thermal  baths interacting through nanoscopic objects? 
Is it possible to accurately control the net heat transfer between nanoparticles and their respective fluctuations by {\em only} applying non-conservative forces, i.e. forces that do not derive from a potential?

Among the broad class of non-conservative forces, nonreciprocal interactions (i.e. forces that violate Newton's third law ``actio=reactio") 
have recently emerged as a topic of lively interest in statistical physics \cite{fruchart2021non,saha2020scalar,you2020nonreciprocity,lavergne2019group,loos2022long}, revealing  nontrivial physical consequences for the dynamical, mechanical and thermodynamic properties of many-body systems. For example, they introduce `odd elasticity' in solids and soft crystals~\cite{braverman2021topological,poncet2022soft} or lead to the formation of travelling waves in binary fluid mixtures~\cite{fruchart2021non,saha2020scalar,you2020nonreciprocity}. In stochastic thermodynamics, recent research has revealed the potential  of nonreciprocal forces in the design of  artificial nano machines  with efficient energetic performance~\cite{Loos2020b}. Inspired by these recent findings, herein we design atomistic MD simulations of trapped nanoparticles immersed in thermal baths at different temperatures that interact through non-conservative, linear forces and that are nonreciprocal. A similar setup was realized experimentally very recently using optical fields \cite{rieser2022tunable}.
Here, we use nonreciprocal interactions to construct a {\em nano refrigerator} which realizes a steady, net heat flow from the cold to the hot bath that does not fulfill Fourier's law for thermal conduction yet  is in  agreement with recent theoretical predictions from ST.  A key advantage of our nano refrigerator design relies on its simplicity as it only requires the usage of nonreciprocal forces acting on each of the nanoparticles. This represents a simplification with respect to previous approaches where heat flows from hot to cold could be achieved using velocity-dependent feedback~\cite{Maxwell1986} or memory registers~\cite{Mandal2012,mandal2013maxwell} as in Maxwell's demons,  or using nonlinear forces in athermal environments~\cite{kanazawa2015minimal}.
 
Our work establishes theoretical design  principles that ensure a prescribed net heat flux in our MD simulations that could be exported to realistic experimental scenarios with trapped nanoparticles~\cite{bechhoefer2005feedback,proesmans2016brownian,ricci2017optically,midtvedt2021fast}. 
We also test  fundamental principles governing the fluctuations of work and the coefficient of performance (COP), some of which follow from recently-established thermodynamic uncertainty relations tested here with realistic atomistic simulations of  nanoparticles~\cite{barato2015thermodynamic,horowitz2020thermodynamic}. These results push forward the synergistic combination of ST  and MD beyond the application of fluctuation theorems in e.g. estimating free energies~~\cite{dellago2014computing,park2003free}.  In particular, our simulations made with  parameters for realistic materials
are a first step towards the engineered design of nanoparticle-based refrigerators powered by thermal fluctuations.

\section{Model and Setup}

\begin{figure}	
\centering
\includegraphics[width=0.4\textwidth]{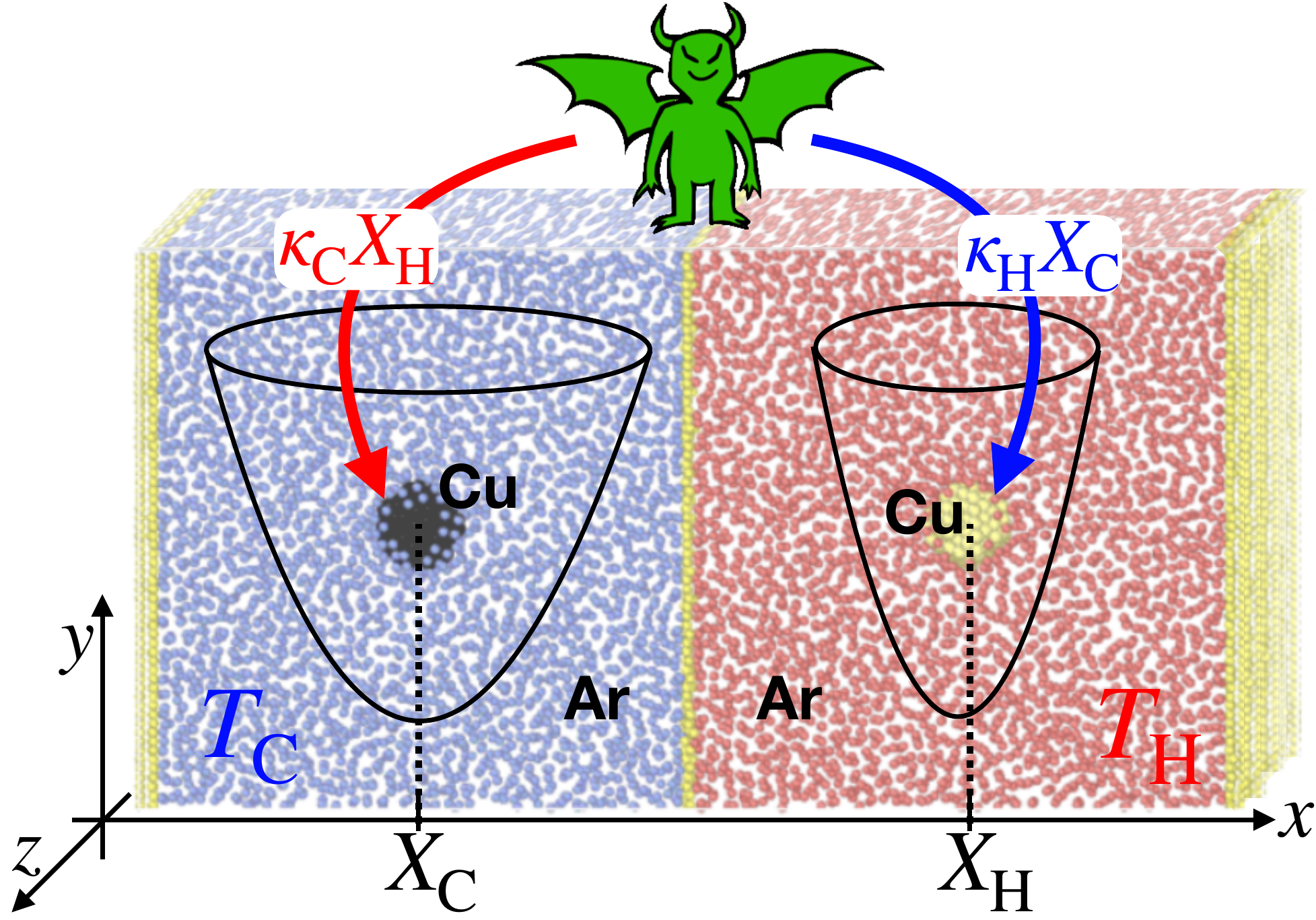}
\caption{
	 Sketch of the molecular dynamics (MD) simulation setup. Two Copper nanoparticles (black and yellow spheres) each consisting of 186 atoms, are immersed in two Argon baths at different temperatures $T_\mathrm{C}=100 \mathrm{K}$ and $T_\mathrm{H}=120\mathrm{K}$. The nanoparticles are trapped  with two static three-dimensional harmonic potentials of different strength (see black line for an illustration). A demon-like controller (green) exerts additional nonreciprocal forces to the two nanoparticles as follows. The demon measures the nanoparticles' positions $X_\mathrm{C},X_\mathrm{H}$ and exerts the forces $\kappa_\mathrm{C}X_\mathrm{H}$ to the nanoparticle in the cold bath and $\kappa_\mathrm{H}X_\mathrm{C}$ to the nanoparticle in the hot bath, where in general $\kappa_\mathrm{H}\neq \kappa_\mathrm{C}$ rendering the demon forces nonreciprocal. 
	}\label{fig:MD}
\end{figure}

%
\subsection{Nanodemon setup and MD simulations}

Constructing an atomistic MD simulation that allows us to violate Newton's third law and furthermore realize a nonreciprocal nano refrigerator, requires a highly unconventional setup in nonequilibrium MD studies. Specifically, we simulate two independent Argon baths that are kept thermostatted at different temperatures ($T_\mathrm{C}=100$K and $ T_\mathrm{H}=120$K). The cold and hot bath are in turn separated by a hard wall made of immobile Copper particles (see Figure 1). Within each bath, we immerse a copper-based spherical nanoparticle of radius $1.4 \mathrm{nm}$.
External nonreciprocal forces (sketched as a green demon in Fig.~\ref{fig:MD}) are applied on the center of each nanoparticle through two  forces, $\kappa_\mathrm{C}X_\mathrm{H}$ and $\kappa_\mathrm{H}X_\mathrm{C}$ on the nanoparticles immersed in the cold and hot baths respectively. Here $X_\mathrm{H}$ and $X_\mathrm{C}$ denote respectively the center-of-mass position of the particle in the hot and the particle in the cold bath, respectively.  Such a setup could, in principle, be realized with the help of an external control scheme (e.g., using optical feedback) similar to~\cite{Lavergne2019,Khadka2018}. As we will see shortly, when $\kappa_\mathrm{C}\neq \kappa_\mathrm{H}$, a \textit{nonreciprocal} coupling is introduced which can lead for specific values $\kappa_\mathrm{C}/\kappa_\mathrm{H}$ to a heat flow from the cold to the hot bath. We have further constrained the particle positions by introducing harmonic potentials (with stiffness $\kappa = 1.16 k_\mathrm{B}T_\mathrm{C}/\AA^2$). The traps prevent the particles from hitting the walls, but are, in principle, not needed to construct the nano refrigerator (i.e., we could also set $\kappa=0$), as is evident from our analytical results introduced in the following.

\subsection{Heat transfer} 

Our MD setup gives us direct access to thermodynamic quantities allowing for quantitative measurements of the heat transfer between the nanoparticles and their respective solvent baths. Specifically, we determined the
total amount of energy change by extracting both the potential and kinetic energy 
of all bath molecules as a function of time which gives the total heat transferred by the copper nanoparticles to both the cold and hot baths, denoted by $\mathrm{d}Q_\mathrm{C}$ and $\mathrm{d}Q_\mathrm{H}$ respectively.
Integrating over the course of the MD simulation yields the $Q_\mathrm{C}$ and $Q_\mathrm{H}$, which directly encodes the stochastic heat dissipated by the nanoparticle into the cold and hot bath respectively. Note that we use the sign convention that 
${Q}>0$ when net energy is dissipated from the nanoparticle to the bath and ${Q}<0$ when it is absorbed by the nanoparticle from the bath.
We estimate the heat dissipation rate $\dot{Q}_\mathrm{C}$ and $\dot{Q}_\mathrm{H}$ from the slope of a  linear regression on the cumulative $Q_\mathrm{C}$ and $Q_\mathrm{H}$ over time. 

\begin{figure}[ht]
\centering
\includegraphics[width=0.49\textwidth]{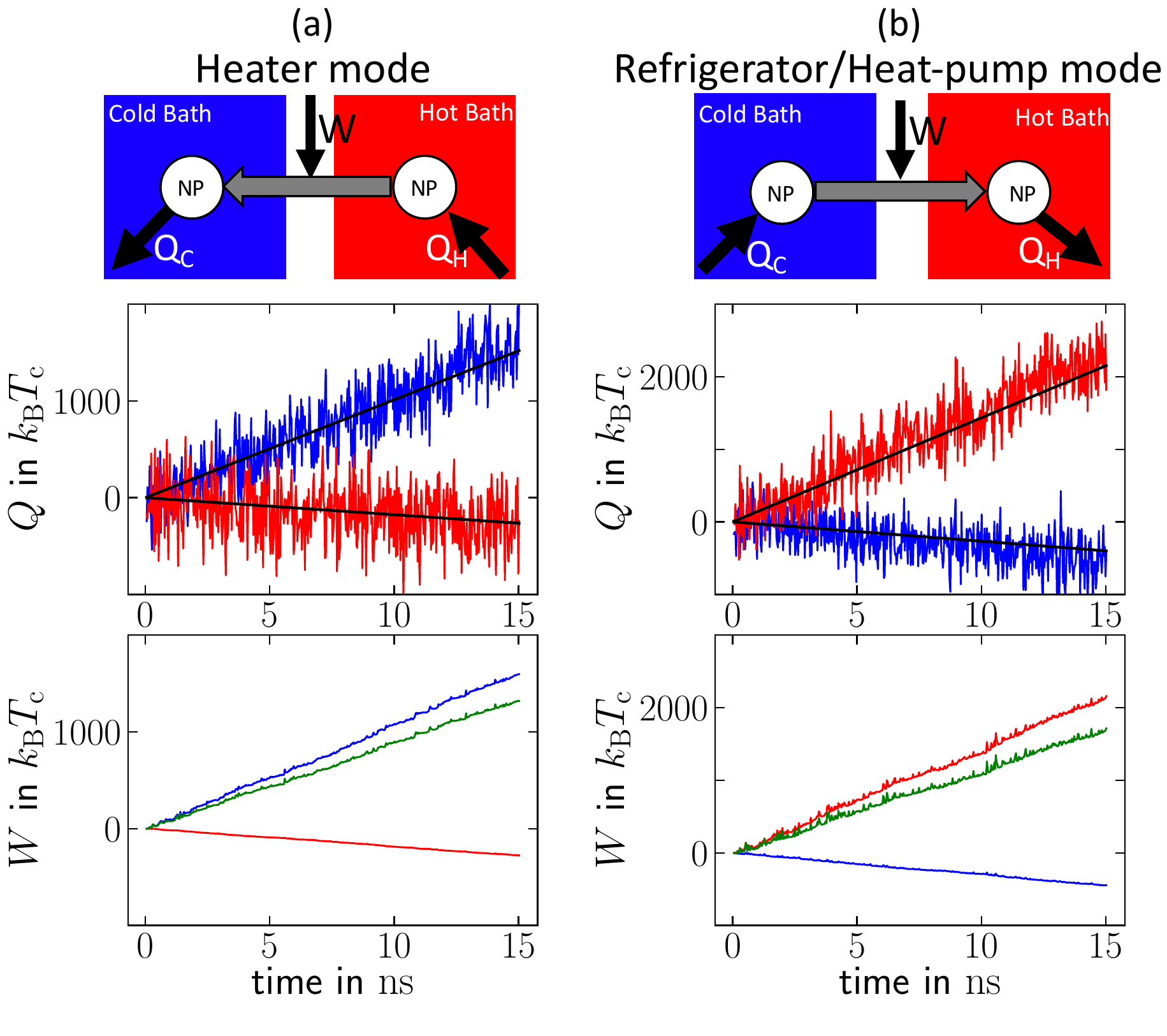}
	\caption{Heat flows $Q$ measured in the MD simulations of the setup sketched in Fig.~\ref{fig:MD}, for two different values of $\kappa_\mathrm{H}$. 
	 (a) {\em Heater Type-II}: for $\kappa_\mathrm{H} =0.17\kappa_\mathrm{C}$,  heat flows from the hot bath to the nanoparticle in the right container and from the nanoparticle to the cold bath in the left container.
	(b)  {\em Refrigerator/Heat-pump}:  for $\kappa_\mathrm{H} =4.80\kappa_\mathrm{C}$,  heat flows from the cold bath to the nanoparticle in the left container and from the nanoparticle to the hot bath in the right container. Here,  heat flows in the reverse direction to the temperature gradient, i.e., it is  extracted from the cold bath and released into the hot bath through the two-nanoparticle system. 	
	\textit{Upper panels} of (a,b) show sketches of the flows of heat and work, respectively. 
	\textit{Middle panels} show the cumulative heat as function of time from the MD simulations, dissipated into the cold bath ($Q_\mathrm{C}$, blue line) and from the hot bath ($Q_\mathrm{H}$, red line).  We use the convention $Q>0$ when heat flows from the nanoparticle into the bath and  $Q<0$  when heat flows from the bath to the nanoparticle. 
	Black lines show linear fits used to obtain the heat rates. \textit{Lower panels} display the stochastic work exerted on the nanoparticle in the cold bath  ($W_{\rm C}$, blue line),  and on the nanoparticle in the hot bath  ($W_{\rm H}$, red line), and the total work given by their sum  ($W=W_{\rm C}+W_{\rm H}$, green line). The work is obtained from the MD simulations by measuring the center of mass positions of the nanoparticles, and evaluating the work using Eq.~\eqref{eq:stochpower}. 
 In both (a, b) we used $\kappa_\mathrm{C}=10\kappa =11.6 k_\mathrm{B}T_\mathrm{C}/\AA ^2 $, $T_\mathrm{C}= 100\mathrm{K}$ and $T_\mathrm{H}=120\mathrm{K}$.} 
\label{fig:heat}
\end{figure}

With this protocol in hand, we begin by demonstrating how tuning the relative strength of $\kappa_\mathrm{C}$ and $\kappa_\mathrm{H}$ provides a microscopic mechanism to alter the direction of heat flow.  Figure~\ref{fig:heat} (a) illustrates $Q_\mathrm{C}$ and $Q_\mathrm{H}$ a situation where the effective coupling force on the 
hot particle is reduced as 
$\kappa_\mathrm{H}\ll \kappa_\mathrm{C}$.
In this case, we observe the canonical situation, where the nanoparticle-duet behaves as a {\em heater}, i.e. heat flows from the hot to the cold bath. On the other hand, by introducing 
an effectively enhanced coupling force experienced by the nanoparticle in the hot bath, there is a striking effect where the direction of the heat flow changes as seen in Figure~\ref{fig:heat} (b)  - heat is now pumped from the cold to the hot bath creating a molecular-scale refrigerator.
The preceding results from the MD simulations provide a powerful proof-of-concept on how an nonreciprocal forces applied on two nanoparticles embedded in  a solvent bath, can in principle be used to change both the rate and direction of heat flow. 

Intuitively, by increasing $\kappa_\mathrm{H}$, the demon tricks the hot particle into \emph{``seeing''} that it was coupled to an even hotter particle, while the cold particle thinks it was coupled to an even colder particle, which results in a heat transfer from the cold to the hot bath. In the following, we will formulate and present a theory that rationalizes this intriguing phenomenon and makes a direct link between thermodynamic observables extracted from the MD simulations and stochastic thermodynamics.

\subsection{Stochastic model}\label{sec:LE}

We employ a mesoscopic stochastic model to describe the  nonequilibrium dynamics of the position and momentum fluctuations of the two-nanoparticle system in their thermal environments. 
To this aim, we describe at a coarse-grained level the dynamics of the $x$-components of the positions and velocities of the center of mass of the nanoparticles, $X_\mathrm{C}$ and $X_\mathrm{H}$, by two coupled underdamped Langevin equations,
\begin{align}\label{eq:Le}
m \ddot{X}_\mathrm{C}+\gamma_\mathrm{C} \dot{X}_\mathrm{C} &=  -(\kappa_\mathrm{C} + {\kappa} ) X_\mathrm{C} + \kappa_\mathrm{C} X_\mathrm{H} + \xi_\mathrm{C}
\\ \label{eq:Le2}
m \ddot{X}_\mathrm{H}+\gamma_\mathrm{H} \dot{X}_\mathrm{H} &
=
\kappa_\mathrm{H}
X_\mathrm{C} -({\kappa}+\kappa_\mathrm{H}) X_\mathrm{H} + \xi_\mathrm{H}.
\end{align}
Here, $m$ is the mass of each nanoparticle, and the coefficients $\kappa_\mathrm{C}$, $\kappa_\mathrm{H}$, $\kappa$ have been defined before. Note that the dynamics is independent of the actual distance between both nanoparticles, which has therefore been excluded from the equations of motions \eqref{eq:Le} by a change of variables (see Methods section).
The stochastic forces $\xi_\mathrm{C}$, $\xi_\mathrm{H}$ are independent Gaussian white noises with zero mean $\langle\xi_\mathrm{C}(t)\rangle=\langle\xi_\mathrm{H}(t)\rangle=0$ modeling the thermal noise exerted by the Argon bath that surrounds  each nanoparticle. 
Their autocorrelation functions are $\langle \xi_j(t)\xi_l(t') \rangle =2 k_\mathrm{B}T_\mu \gamma_\mu \,\delta_{jl}\delta(t-t')$, where $ j,l\in \{\mathrm{C,H}\}$ are indices denoting the hot or cold bath, $\delta_{jl}$ is Kroneckers' delta, and $k_\mathrm{B}$ is  Boltzmann's constant. Here and in the following, $\langle . \rangle$ denote averages over many realizations of the noise. The averages obtained from the MD simulations are extracted from {single} trajectories of $15$ns long, i.e. exceeding by three orders of magnitude the relaxation times of our system (see below). %
Furthermore, $\gamma_\mathrm{C}$ and $\gamma_\mathrm{H}$ are {\em effective}  coefficients of the friction  that each nanoparticle experiences in its respective environment.
Despite its simplicity, the model \eqref{eq:Le} allows us to infer dynamical and thermodynamic properties of our molecular dynamics simulations, as we describe below.

An important ingredient for the theory are the effective friction coefficients $\gamma_\mathrm{C}$ and $\gamma_\mathrm{H}$, which  emerge from the interaction of the nanoparticles with the baths' particles~\cite{stewart1962transport}. To estimate  $\gamma_\mathrm{C}$ and $\gamma_\mathrm{H}$, we run an equilibrium simulation in the absence of (nonreciprocal) interactions by taking $\kappa_\mathrm{C}=\kappa_\mathrm{H}=0$. In this limit, the position and velocity autocorrelation functions can be solved analytically (see Methods section). By fitting the autocorrelation functions obtained from the MD simulations to the analytical formulas, we extract the estimates $\gamma_\mathrm{C}\simeq \gamma_\mathrm{H}  \sim 3.5 \times 10^{-12}$kg/s. This value is consistent with an independent estimate obtained from a \emph{dragging experiment} simulation where a nanoparticle is pulled with a constant force along  the $x$ axis (see SI for further details), which yields an estimate of $\gamma_\mathrm{C} \approx \gamma_\mathrm{H} \approx 3\times 10^{-12}$kg/s. In the following and for further analyses, we use the estimate $\gamma = 3.5 \times 10^{-12}$kg/s for the friction of the two nanoparticles. 

In the ensuing analysis, several timescales are relevant to characterize the dynamics of the particles at the mesoscopic level, following~\eqref{eq:Le} and~\eqref{eq:Le2}.
Firstly, the momentum relaxation time extracted from our data $m/\gamma \approx 5$ps is only one order of magnitude smaller than the relaxation time for the position in the trap $\gamma/{\kappa} \approx 25$ps. %
These results lend credence to the validity of the underdamped description used in our approach, as the  data acquisition timescale $\Delta t\sim 1$ps is below the
momentum relaxation timescale.


\subsection{Stochastic Energetics\label{sec:Energy}} 

\begin{figure}[t]
\centering
\includegraphics[width=0.48\textwidth]{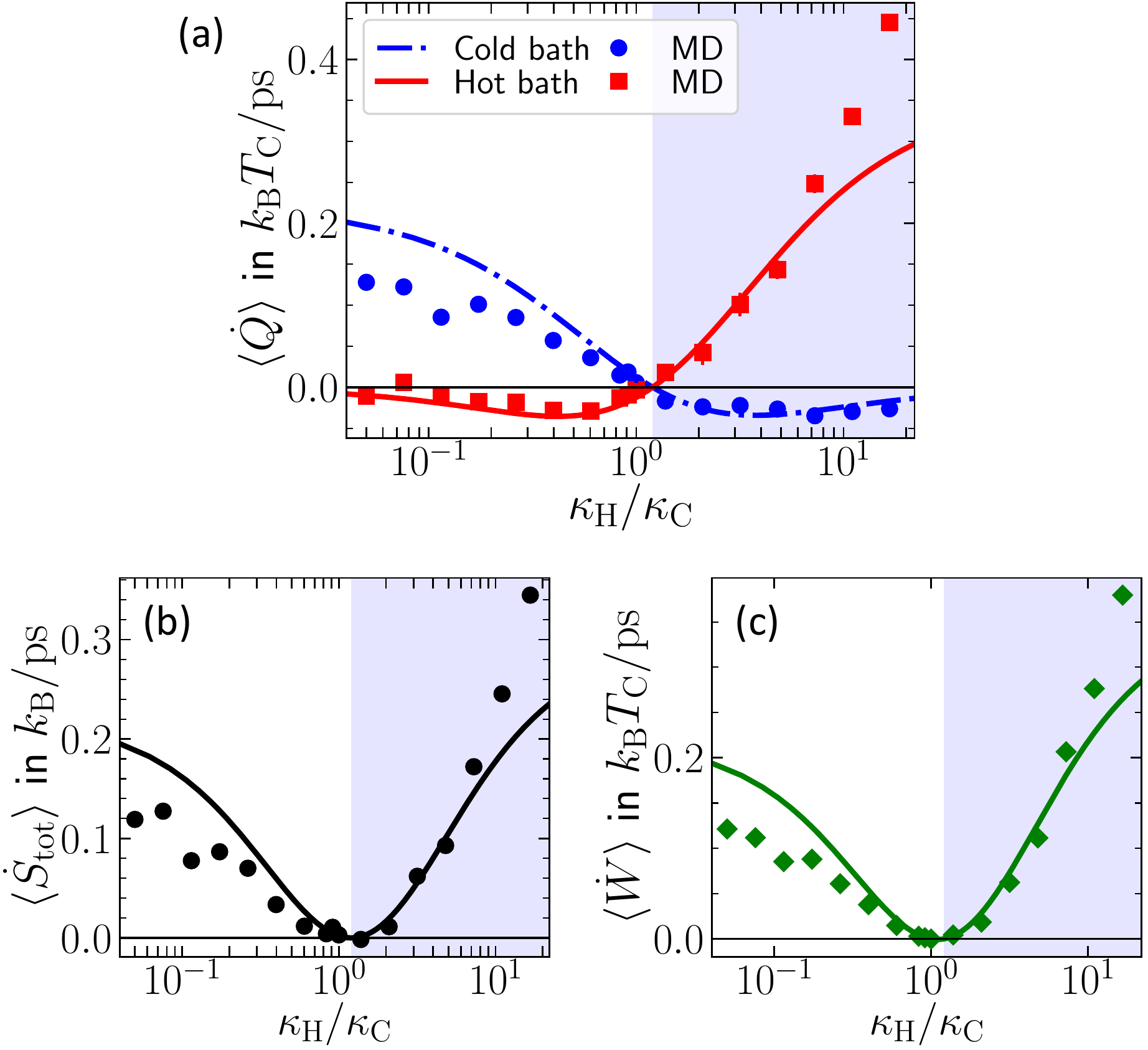}\\
	\caption{Comparison between thermodynamic fluxes obtained from MD simulations (symbols) and theoretical predictions from stochastic thermodynamics (lines): 	 The average rate of  heat dissipation rate in each bath (a), the average entropy production rate (b), and the total power applied by the demon to the system (c), as a function of the coupling constant $\kappa_\mathrm{H}$. 
   The heat rates  in (a) are measured by accumulating the amount of energy extracted by the thermostats and linearly fitting the results to obtain the slope, see Fig.~\ref{fig:heat}. 
The blue shaded area illustrates the parameter regime at which the system works as a refrigerator (given by  the condition~\ref{cond1}).}\label{fig:heat2}
\end{figure}

We now develop and put to the test a theory for the energetics of the nanoparticle system setup in the light of stochastic thermodynamics~\cite{Sekimoto2010}---a framework that enables to describe the fluctuations of heat and work of systems described by e.g. Langevin equations, such as  Eq.~(\ref{eq:Le}).
Applying the framework of stochastic thermodynamics to the model given by  Eq.~(\ref{eq:Le}), and putting forward  mathematical techniques introduced in~\cite{kwon2005structure,bae2021inertial,kwon2011nonequilibrium},  we derive exact closed-form expressions for the expected value of the rate of heat dissipated into the cold and  hot baths' in the steady state, reading as,
\begin{eqnarray}
 \langle\dot{Q}_\mathrm{C}\rangle
 =-k_\mathrm{B}
\frac{ \displaystyle \kappa_\mathrm{C} (\kappa_\mathrm{C} T_\mathrm{H}-\kappa_\mathrm{H} T_\mathrm{C})}{\displaystyle(m/\gamma)  (\kappa_\mathrm{C} + \kappa_\mathrm{H})^2 /2+ \gamma  (2\kappa+\kappa_\mathrm{C}+\kappa_\mathrm{H})} ,
 \label{eq:Qc}
\\
 \langle\dot{Q}_\mathrm{H}\rangle
 =-k_\mathrm{B}
\frac{\displaystyle- \kappa_\mathrm{H} (\kappa_\mathrm{C} T_\mathrm{H}-\kappa_\mathrm{H} T_\mathrm{C})}{\displaystyle(m/\gamma)  (\kappa_\mathrm{C} + \kappa_\mathrm{H})^2 /2+ \gamma  (2\kappa+\kappa_\mathrm{C}+\kappa_\mathrm{H})} . \label{eq:Qh}
\end{eqnarray}
To ensure that the dynamics of the nanoparticles is stable, we further find the necessary condition $\kappa_\mathrm{H} + \kappa_\mathrm{C} > -\kappa$ (see Methods section).
Note that in the above formulas, the steady-state averages $\langle\dot{Q}_\mathrm{C}\rangle= \langle \mathrm{d}{Q}_\mathrm{C}/\mathrm{d}t\rangle$ and $\langle\dot{Q}_\mathrm{H}\rangle= \langle \mathrm{d}{Q}_\mathrm{H}/\mathrm{d}t\rangle$ are obtained using the definitions of stochastic heat used in stochastic thermodynamics  (see Methods section), which are written in terms of the nanoparticles' positions  and velocities and thus not necessarily equal to the ``direct" stochastic heat measured in the MD simulations from the energy fluctuations of the bath molecules. 

Notably, Eqs.~(\ref{eq:Qc}-\ref{eq:Qh}) predict a  net heat transfer between the two baths that obeys Fourier's law $\langle\dot{Q}_\mathrm{C,H}\rangle \propto (T_\mathrm{H}-T_\mathrm{C})$
only when the forces exerted by the demon are reciprocal. Moreover,  
our theory predicts that a net heat flow can be induced by three mechanisms: first, the existence of a temperature gradient, second, nonreciprocal coupling, or, third, a combination of both. 
As we show below, such a net heat flow is a signature of entropy production, with the latter being also accessible from our theory. In particular,  Eqs.~(\ref{eq:Qc},\ref{eq:Qh}) allow for the prediction of a closed-form theoretical expression for the steady-state average rate of entropy production $
\langle \dot{S}_\mathrm{tot}\rangle = 
\langle \dot{Q}_\mathrm{H}\rangle /T_\mathrm{H} +\langle \dot{Q}_\mathrm{C} \rangle/T_\mathrm{C}$~\cite{Seifert2012}, yielding
\begin{align}
\langle \dot{S}_\mathrm{tot} \rangle
& =
\frac{k_\mathrm{B}( \kappa_\mathrm{C}T_\mathrm{H}  - \kappa_\mathrm{H} T_\mathrm{C})^2/(T_\mathrm{C}T_\mathrm{H})}
{(m/\gamma) (\kappa_\mathrm{C} +\kappa_\mathrm{H})^2/2 + \gamma(\kappa_\mathrm{C}+\kappa_\mathrm{H}+2\kappa)}
.\label{eq:Stot} 
\end{align}
Note that $\langle \dot{S}_\mathrm{tot} \rangle\geq 0$ for any parameter values, in agreement with the second law of stochastic thermodynamics, with equality only for the choice $\kappa_\mathrm{H}/ \kappa_\mathrm{C}= T_\mathrm{H}/T_\mathrm{C}$ for the nonreciprocal coupling constants. Furthermore, from Eqs.~(\ref{eq:Qc}-\ref{eq:Qh}) we can also extract the total power exerted by the demon on the nanoparticles, which is simply given by $\langle \dot{W}\rangle = \langle \dot{Q}_\mathrm{C}+\dot{Q}_\mathrm{H} \rangle$, see below.

To gain further insights, we compare  in Fig.~\ref{fig:heat2}a our theoretical predictions from stochastic thermodynamics (Eqs.~(\ref{eq:Qc}-\ref{eq:Qh}), lines) for the heat transfer evaluated directly over a collection of MD simulations (symbols) ran over a wide range of values for the demon coupling constants $\kappa_\mathrm{H}$ and $\kappa_\mathrm{C}$ spanning three orders of magnitude in $\kappa_\mathrm{H}/\kappa_\mathrm{C}$. Figure ~\ref{fig:heat2}a reveals an excellent semi-quantitative agreement between the direct heat measurement in MD simulations with the stochastic theory over the  parameter regime that we explore.  Notably, we remark that  our predictions  are done   without using any fitting parameter, i.e. we use in Eqs.~(\ref{eq:Qc}-\ref{eq:Qh}) the actual parameter values of the MD simulation and the friction coefficient estimated from the equilibrium fluctuations described above. Importantly, the MD simulation results reveal that the system acts as a heater whenever $\kappa_\mathrm{H}/ \kappa_\mathrm{C}< T_\mathrm{H}/T_\mathrm{C}$ and as a refrigerator (and heat-pump) when 
\begin{equation}\label{cond1}
\kappa_\mathrm{H}/ \kappa_\mathrm{C}> T_\mathrm{H}/T_\mathrm{C}, 
\end{equation}
a feature that is supported by our theory.  This result rationalizes  the findings in Fig.~\ref{fig:heat}, which correspond to $\kappa_\mathrm{H}/ \kappa_\mathrm{C}=0.17$ (Fig.~\ref{fig:heat}a) and $\kappa_\mathrm{H}/ \kappa_\mathrm{C}=4.8$ (Fig.~\ref{fig:heat}b) which correspond respectively to the heater and refrigerator regimes as $T_\mathrm{H}/T_\mathrm{C}=1.2$.  

We also compare in Fig.~\ref{fig:heat2}b  the theoretical prediction~\eqref{eq:Stot} for steady-state rate of entropy production (line) with its value  estimated from the MD simulations (symbols), with the latter evaluated by plugging in to $\langle \dot{S}_\mathrm{tot}\rangle = 
\langle \dot{Q}_\mathrm{H}\rangle /T_\mathrm{H} +\langle \dot{Q}_\mathrm{C} \rangle/T_\mathrm{C}$, the MD values of the heat flows in the thermostats divided by their respective temperatures. 

The estimate for the rate of entropy production obtained from the MD measurements is 
in excellent agreement with the  theoretical expression given by~\eqref{eq:Stot}.  Around the  coupling values $\kappa_\mathrm{C} T_\mathrm{H} =\kappa_\mathrm{H} T_\mathrm{C}$ the heat flow and entropy production vanish, which we will refer to as a ``pseudo equilibrium'' point in the following.  In the SI we show that in this case, detailed balance holds, i.e., all probability currents vanish, despite the presence of a temperature gradient in the system together with a nonreciprocal coupling.

\section{Power and Performance}
\begin{figure}[t]
\centering
\includegraphics[width=0.28\textwidth]{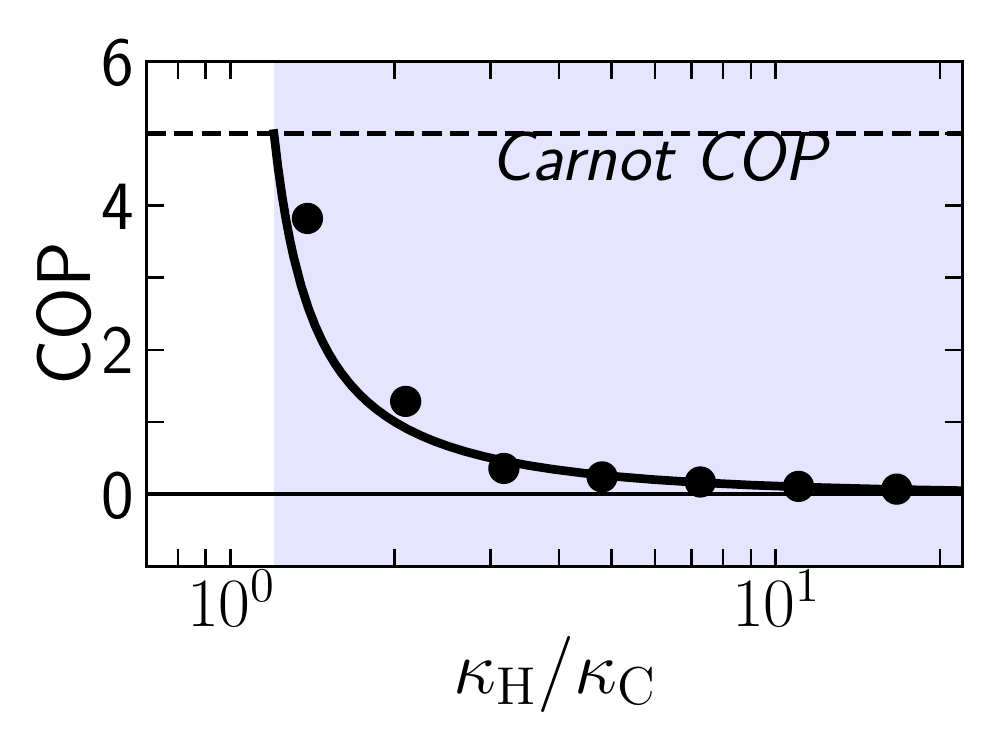}
	\caption{ Average coefficient of performance as a function of $\kappa_\mathrm{H}$  in the refrigeration mode: results from MD simulations (symbols) and theoretical prediction from stochastic thermodynamics [\eqref{eq:COP_r}, line]. 
	The horizontal line indicates the Carnot bound for the coefficient of performance, reached at $\kappa_\mathrm{H}= \kappa_\mathrm{C} T_\mathrm{H}/T_\mathrm{C}$, and given by   $ {\rm{COP}}  = T_\mathrm{C}/ (T_\mathrm{H} -T_\mathrm{C})$.
	The rest of the simulation parameters were set the same as in Fig.~\ref{fig:heat}.
	}
	\label{fig:COP}
\end{figure}
\subsection{Performance of the nano refrigerator}

In any refrigerator, getting the most heat from the temperature source is desirable by doing the least amount of work possible. Therefore, a suitable coefficient of performance (COP) is defined as the ratio of the heat taken from the low-temperature source to work done on the machine. A higher COP indicates a better economic performance of the system. In macroscopic systems, COP is usually in the range of $1-4$~\cite{borgnakke2022fundamentals}.
To quantify the net energetic  performance of the nano refrigerator,  we evaluate its coefficient of performance, defined by the average rate of heat that is extracted from the cold bath 
divided by the average total power inputted into the system, COP$=|\langle \dot{Q}_\mathrm{C}\rangle |/\langle \dot{W}\rangle $.  
From this definition and upon using  our theoretical  predictions for the heat transfer  given by  Eqs.~(\ref{eq:Qc},\ref{eq:Qh}), we predict that the COP follows
\begin{align}
\rm{COP}
\equiv 
\frac{|\langle \dot{Q}_\mathrm{C}\rangle |}{\langle\dot {W}\rangle} 
=
\frac{-\langle \dot{Q}_\mathrm{C}\rangle }{\langle\dot{Q}_\mathrm{H}\rangle+\langle\dot{Q}_\mathrm{C}\rangle}
 =
\frac{\kappa_\mathrm{C}}
{\kappa_\mathrm{H}-\kappa_\mathrm{C}} .
\label{eq:COP_r}
\end{align}
In \eqref{eq:COP_r}, the second equality follows from using the first law of thermodynamics for stationary states which in this case reads  $\langle\dot{W}\rangle-\langle \dot{Q}_\mathrm{C}\rangle-\langle \dot{Q}_\mathrm{H}\rangle=0$, whereas the third equality follows from   Eqs.~(\ref{eq:Qc},\ref{eq:Qh}). 
Remarkably, our theoretical prediction for the COP   depends solely on the nonreciprocal coupling parameters $\kappa_\mathrm{H}$ and $\kappa_\mathrm{C}$.
Figure~\ref{fig:COP}  reveals a good agreement between the value of the $\mathrm{COP}$ estimated from the MD simulations (symbols) and the prediction from our stochastic theory  \eqref{eq:COP_r} (lines).  %
Interestingly the  agreement between simulation and theory is enhanced especially in far from equilibrium conditions, i.e.  for large  relative strengths of nonreciprocity $\kappa_\mathrm{H}/\kappa_\mathrm{C}\gg 1$. Close to the pseudo equilibrium $\kappa_\mathrm{H}/\kappa_\mathrm{C}=T_\mathrm{H}/T_\mathrm{C}$ point, the noise in the simulation results seems more pronounced. Towards this point, the theory predicts Carnot efficiency, which corresponds to a COP of $T_\mathrm{C}/(T_\mathrm{H}-T_\mathrm{C})=5$ in the present case.

\subsection{Fluctuations of power and performance}

We have shown so far that,  when $\kappa_\mathrm{H}/\kappa_\mathrm{C}>T_\mathrm{H}/T_\mathrm{C}$,  the nanoparticle system behaves like a refrigerator \textit{on average}, i.e. its steady-state average fluxes obey $\langle  \dot{Q}_\mathrm{C}\rangle>0 $, $\langle  \dot{Q}_\mathrm{H}\rangle<0 $ and $\langle  \dot{W}\rangle>0$.  Due to thermal fluctuations, the two-nanoparticle system can give rise to  transient values of the fluxes that do not obey the refrigerator constraints (e.g. transient values  $ \dot{Q}_\mathrm{C}<0 $, $  \dot{Q}_\mathrm{H}>0 $ and $  \dot{W}<0$ in a small time interval) as revealed in Langevin-dynamics models~\cite{rana2016anomalous}. In order to inspect such  fluctuation phenomena it is mandatory to evaluate  quantities such as the  power and  performance of the nano machine  along individual, short time intervals. 
As a first approach in this direction, we evaluate the  fluctuations of the stochastic power from the MD simulations using the
positional fluctuations of the center of mass of the nanoparticles. In particular, we evaluate the stochastic power exerted in a small time interval $[t,t+\mathrm{d}t]$ using the expression from stochastic thermodynamics~\cite{Sekimoto2010} associated with the model given by~\eqref{eq:Le}
\begin{equation}\label{eq:stochpower}
\dot W  = \underbrace{\kappa_\mathrm{C} X_\mathrm{H} \circ \dot{X}_\mathrm{C}}_{=\displaystyle \dot{W}_\mathrm{C}} +  \underbrace{\kappa_\mathrm{H} X_\mathrm{C}\circ  \dot{X}_\mathrm{H}}_{=\displaystyle \dot{W}_\mathrm{H}}.
\end{equation}
Here, $\circ$  the Stratonovich product, whereas $ \dot{X}_\mathrm{C} =[X_\mathrm{C} (t+\mathrm{d}t) - X_\mathrm{C} (t)]/\mathrm{d}t$ and $ \dot{X}_\mathrm{H} =[X_\mathrm{H} (t+\mathrm{d}t) - X_\mathrm{H} (t)]/\mathrm{d}t$  are the  time-averaged velocities estimated from the positions of the nanoparticles. We have also introduced  $ \dot{W}_\mathrm{C}$ and $\dot{W}_\mathrm{H}$ as  the fluctuating power exerted on the cold (hot) nanoparticle in the interval $[t,t+\mathrm{d}t]$ respectively.  Note that in~\eqref{eq:stochpower} we take into account only the non-conservative forces exerted on each nanoparticle due to their nonreciprocal coupling. 

\begin{figure}
    \centering
    \includegraphics[width=0.48\textwidth]{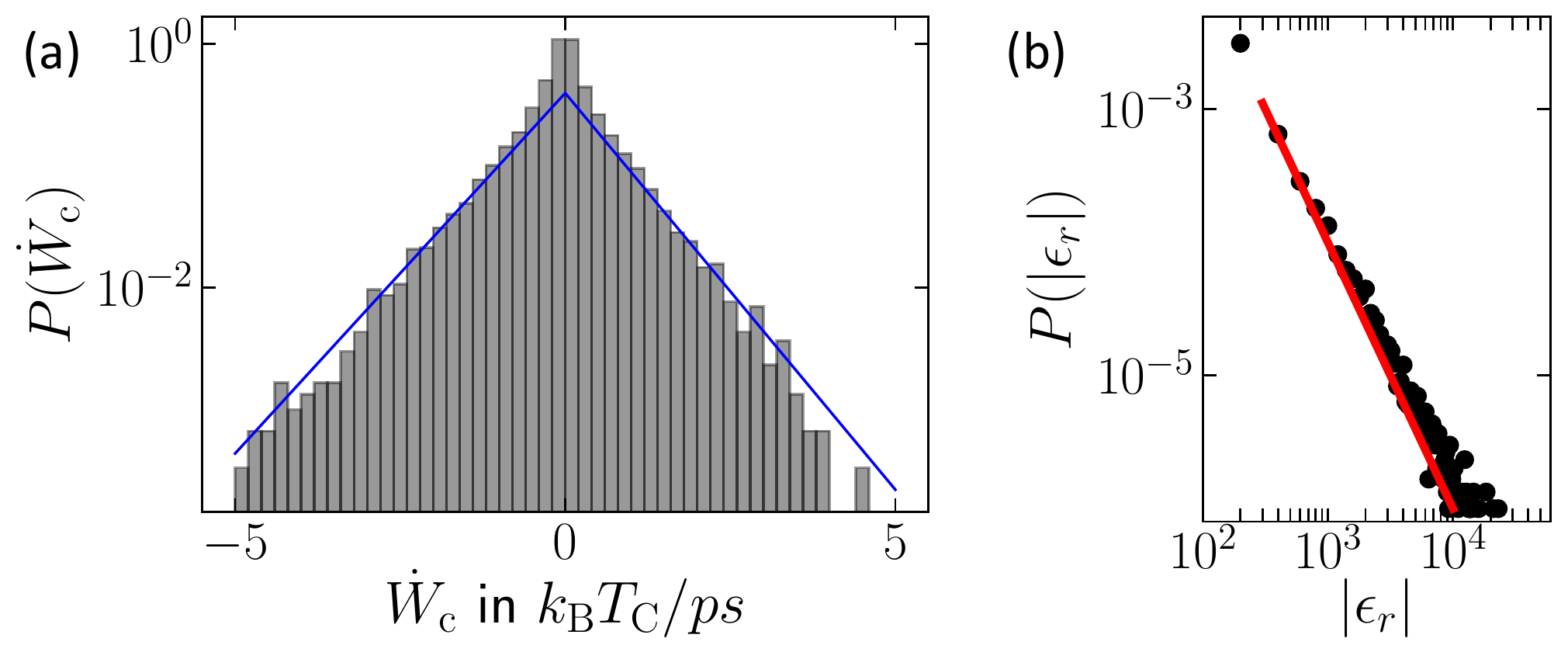}
    \caption{
    (a) Distribution of the stochastic power exerted on the cold nanoparticle: result from the MD simulations (bars) and  analytical prediction given by~\eqref{eq:Pw} (solid blue line). Note that the lower panel in Figure 2 (b) displays the corresponding values of $W_\mathrm{C}$.
    (b) Distribution of the stochastic coefficient of performance defined by~\eqref{def:cop}: empirical density obtained  from the MD simulations
     (see Methods section) and theoretical prediction given by a   power law with exponent $-2$ (red line). For the second panel we took $\kappa_\mathrm{H}/\kappa_\mathrm{C}=7.27$ and fixed the rest of the parameter values the same as in Fig.~\ref{fig:heat2}.
    }
    \label{fig:power+efficiency_fluctuations}
\end{figure}

We now evaluate the stochastic power~\eqref{eq:stochpower} from our MD simulations over time intervals of duration $\mathrm{d}t=1$ps, for the parameter value  $\kappa_\mathrm{H}/\kappa_\mathrm{C}=4.8$ where we obtained a maximum heat  extraction on average from the cold bath. To this aim, we extract the empirical probability density for the  stochastic power  which reveals considerable fluctuations  (gray bars in Fig.~\ref{fig:power+efficiency_fluctuations}a). The distribution estimated from the MD simulations can be described with impressive accuracy using the  closed-form expressions  for the distribution of  $\dot W_\mathrm{C}$ and $\dot W_\mathrm{H}$
\begin{equation}\label{eq:Pw}
P(\dot W_j)= 
\frac{1}{\mathcal{Z}_j} \,\exp \left [
 \frac{\beta_j}{\kappa_j}  \dot{W}_j  
 -2\frac{\sqrt{\zeta_j \alpha_j}}{|\kappa_j|} |\dot{W}_j| \right], 
\end{equation}
which  we derive analytically using the definition of the stochastic power~\eqref{eq:stochpower} and assuming the  effective stochastic model~\eqref{eq:Le}. In~\eqref{eq:Pw}, $j\in\{\mathrm{C,H}\}$ is the particle label,   $\mathcal{Z}_j$ is a normalization constant, and $\alpha_j, \beta_j, \zeta_j$ are functions of $\kappa_\mathrm{C},\kappa_\mathrm{H},T_\mathrm{C},T_\mathrm{H},\gamma,m$ (see Methods section for their explicit expressions).  
The power distributions 
have exponential tails and are slightly asymmetric. In the refrigerator mode, $P(\dot W_\mathrm{C})$ leans towards positive values, consistent with the net negative work value, while $P(\dot W_\mathrm{H})$ leans towards positive work values. 
Furthermore, we find a good agreement between the MD simulation results and \eqref{eq:Pw}  throughout the parameter regime that we explore. The theoretical predictions yield a systematic  overestimation of the variance  which we attribute to the finite timestep $\mathrm{d}t$ in the MD simulations, while in the theoretical calculations we assume it to be infinitesimal (see Supplemental Material for  the values of the variance of the power for different values of $\kappa_\mathrm{H}/\kappa_\mathrm{C}$).

The previous analyses showing that the work and heat in a small time interval are highly fluctuating motivates us to investigate the finite-time fluctuations of the coefficient of performance of the nano machine. To quantify how much the efficiency fluctuates for individual trajectories, we consider the {\em stochastic} coefficient of performance defined as~\cite{joseph2021efficiency}
\begin{align}\label{def:cop}
    \epsilon_r =& \frac{|\dot{Q}_\mathrm{C}|}{\dot{W}}.
\end{align}
Note that since in general  $\langle|\dot{Q}_\mathrm{C}|\rangle/\langle\dot W\rangle\neq \langle|\dot{Q}_\mathrm{C}|/\dot{W}\rangle  $, the ensemble average of $\epsilon_r$ does not coincide  with the average COP given by~\eqref{eq:COP_r}. 
Figure~\ref{fig:power+efficiency_fluctuations}b displays the empirical distribution of the stochastic COP obtained  from the MD simulations, which develops a fat tail with values that can exceed significantly  the Carnot value~$\epsilon_\mathrm{C}   =T_\mathrm{C}/ (T_\mathrm{H}-T_\mathrm{C})=5$. Such super-Carnot performance achieved in short time intervals have also been reported in experimental conditions and theoretical models of nanoscopic heat engines ~\cite{polettini2015efficiency,verley2014unlikely,martinez2017colloidal,park2016efficiency} and  nano refrigerators~\cite{rana2016anomalous}; they result from  rare events where the refrigerator works transiently as a heater  reversing the work flux with respect to its average behaviour.
Furthermore we find that the distribution of $\epsilon_r$ extracted form the MD simulations  follows in good approximation a power law 
\begin{equation}
    P(\epsilon_r)  \simeq  |\epsilon_r|^{-2},
    \label{eq:pl}
\end{equation}
see red line in Fig.~\ref{fig:power+efficiency_fluctuations}b. 
The power-law behavior that we find reinforces the critical  significance of thermal fluctuations for our nano machine system setup. Remarkably, we find that the power law~\eqref{eq:pl} is in good agreement with the numerical results for all values of $\kappa_\mathrm{H}$ and $\kappa_\mathrm{C}$ that we explored, suggesting a universal scaling behavior as predicted by previous theoretical  work within the realm of stochastic efficiency~\cite{polettini2015efficiency}.

\subsection{Uncertainty relations}

The results in previous sections revealed the instrumental role of stochastic thermodynamics to establish design principles for the parameter values of the nano machine to achieve prescribed values of the net power and efficiency.  
In the following, we investigate how one can use principles from stochastic thermodynamics --namely the so-called thermodynamic uncertainty relations~\cite{barato2015thermodynamic,hasegawa2019uncertainty,horowitz2020thermodynamic}--  to put fundamental constraints  that regulate the trade-off between dissipation and precision of the nano machine.

\begin{figure}
    \centering
    \includegraphics[width=0.4\textwidth]{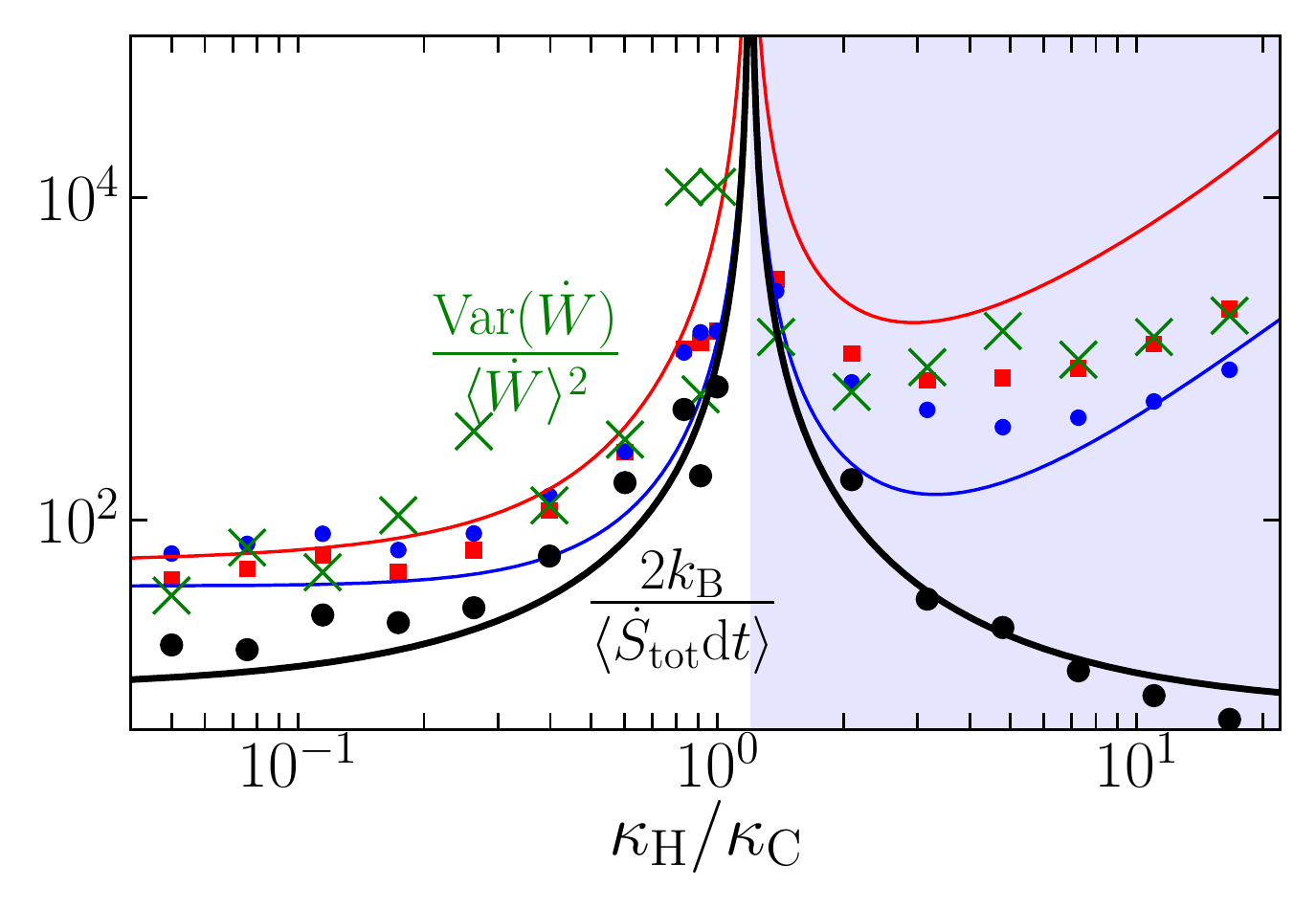}
    \caption{
    Thermodynamic uncertainty relation for power fluctuations. Relative uncertainty $\text{Var}(\dot{W})/\langle \dot{W}\rangle^2$ of the stochastic power exerted on the cold $\dot{W}_\mathrm{C}$ (blue filled circles) and the hot $\dot{W}_\mathrm{H}$  (red filled squares) nanoparticles, computed using in Eq.~\eqref{eq:stochpower}  the output of our MD simulations. The uncertainty of the total power on the two-nanoparticle setup $\dot{W}=\dot{W}_\mathrm{C}+\dot{W}_\mathrm{H}$ is show by green crosses. The blue and red lines are given by our theoretical predictions for the relative uncertainty of $\dot{W}_\mathrm{C}$ and $\dot{W}_\mathrm{H}$, respectively (see Supplemental for the full derivation). 
    The relative uncertainty of the stochastic power lies above  2 divided by   the average   entropy production  during ${\rm d}t$ [black symbols, MD simulations; black line, theoretical prediction given by \eqref{eq:Stot}], in agreement with the thermodynamic uncertainty relation  
    \eqref{eq:TUR}. 
    }
    \label{fig:power_uncertainty}
\end{figure}

A suitable measure to quantify the strength of the power fluctuations is the \textit{uncertainty} of the power, defined by the variance over the squared mean. High values of uncertainty indicate that the dynamics is essentially dominated by  fluctuations. 
We have measured the uncertainty of the total power $\dot W = \dot W_\mathrm{C} + \dot W_\mathrm{H}$ in the MD simulations from individual trajectories, by making a statistics over the extracted values of the work. Figure \ref{fig:power_uncertainty} shows the results for the uncertainty of the total power (green symbols), as well as the power uncertainties of $\dot{W}_\mathrm{C}$ and $\dot{W}_\mathrm{H}$ separately (blue and the red symbols). Remarkably, the uncertainties reach extremely high values around the pseudo equilibrium point, and even seem to diverge at $\kappa_\mathrm{H} /\kappa_\mathrm{C} = T_\mathrm{H}/T_\mathrm{C}$. 
As we show below, this blow-up can be understood by making use of a recently-developed trade-off relation between the precision of thermodynamic currents and the rate of entropy production. 

In the field of stochastic thermodynamics, recently a universally class of results ---often called e thermodynamic uncertainty relation--- governing Markovian nonequilibrium stationary states  were derived, see e.g.~\cite{barato2015thermodynamic,hasegawa2019uncertainty,horowitz2020thermodynamic}.
Such laws connect the uncertainty of a current, quantified by its signal-to-noise ratio, with the total thermodynamic cost, measured by the steady-state rate of entropy production. In particular, the thermodynamic uncertainty relation in Ref.~\cite{pietzonka2017finite} implies  that the uncertainty of the finite-time power fluctuations of stationary Markovian processes is always bounded from below by $2k_\mathrm{B}$ over the mean total entropy production during ${\rm d}t$, i.e., 
\begin{equation}\label{eq:TUR}
    \frac{\mathrm{Var}({\dot{W}})}{\langle \dot W \rangle^2} \geq \frac{2 k_{\rm B}}{\langle  \dot S_\mathrm{tot}\rangle {\rm d}t}.
\end{equation} 
Equation~\eqref{eq:TUR} reveals that there exists  a minimal thermodynamic cost associated with  achieving a certain precision of the power exerted by the controller.  
Applying this law to the present case, we find from \eqref{eq:Stot} that $\langle \dot S_\mathrm{tot} \rangle \to 0$ at $\kappa_\mathrm{H} /\kappa_\mathrm{C} \to T_\mathrm{H}/T_\mathrm{C}$, thus, the power uncertainty indeed diverges at the pseudo equilibrium point.
We have complemented the MD simulation results in Fig.~\ref{fig:power_uncertainty} with a black line showing the lower bound $2 k_{\rm B}/\langle  \dot S_\mathrm{tot}\rangle{\rm d}t$ according to~\eqref{eq:TUR}, which provides the first test of thermodynamic uncertainty relations in nonequilibrium MD setups.

From the simulation results shown in Fig. \ref{fig:power_uncertainty}, we further detect a minimum in the total power uncertainty in the refrigerator regime. Interestingly, the position of this minimum (that is $\kappa_\mathrm{H}/\kappa_\mathrm{C} \approx 2$) roughly coincides with the  $\kappa_\mathrm{H}/\kappa_\mathrm{C}$ value where the amount of extracted heat is maximal. Thus, there is a regime where the refrigeration is maximal while at the same time, the power to sustain the refrigerator is as precise as possible.


\section{Discussion}

We have shown with molecular dynamics simulations that it is possible to attain a net heat flow from a cold to a hot bath by connecting two  nanoparticles immersed in fluid containers through nonreciprocal forces. Such nonreciprocal forces could be realized in the laboratory upon using, e.g., feedback traps~\cite{Jun2012} which allow to exert in real time forces  based on measurements of only the position of the center of mass of the nanoparticles. This represents an advantage with respect to traditional Maxwell-demon approaches where the measurement of position and velocities of the bath molecules is required to attain a heat flow from a cold to a hot thermal bath. Notably, our setup  is also advantageous with respect to previous theoretical proposals that require the usage of athermal fluctuations together with nonlinear forces between the nanoparticles~\cite{Kanazawa2013}.  

In our simulations we have studied the case of silver nanoparticles immersed in argon, however we expect this effect to be generic for a class of systems whose dynamics can be described by linear underdamped Langevin equations. For such class of systems, we have revealed by using stochastic thermodynamics the necessary conditions (i.e., design principles) that ensure a reverse heat flow (from cold to hot) and provided predictions on key thermodynamic properties such as the net heat transfer and the coefficient of performance. Our theoretical results reveal that the refrigeration effect generated by the nonreciprocal coupling is robust. It can be achieved in a broad parameter range as long as the dynamics is stable (see Appendix). 

This work demonstrates fruits of the bridge between molecular dynamics with stochastic thermodynamics, namely the possibility to establish quantitative criteria to control the statistics of thermodynamic fluxes in nanoparticle-based thermal machines. We have developed analytical formulae that describe the average heat fluxes between the nanoparticles, the entropy production rate and the coefficient of performance. Moreover, we have tackled analytically the statistics of the power and the coefficient of performance of the nano machine over  short time intervals within fluctuations play a prominent role. From the theoretical perspective, we expect that stochastic-thermodynamic approaches could shed light in the future on optimal design of nano refrigerator devices taking into account e.g. the effect of delay and/or memory in the exertion of nonreciprocal forces that may play a key role in realistic experimental scenarios.

In the current era when efficient energy conversion is of paramount importance,  optimizing thermal transport in both classical and quantum systems has become a topic of intense theoretical and experimental interest for nanotechnology~\cite{baroni2017,baroni2019}. Furthermore, the possibility of realizing thermal conduits in various biological systems have been proposed to optimize thermal networks and information transfer~\cite{klinman2021,klinman2022}. Transient reverse heat flow in these contexts may confer biological machinery with enhanced functionality for energy harvesting. In this regard, it would be interesting to theoretically investigate whether our theoretical and atomistic setup can be used to generate heat flows in particles immersed in more complex environments including viscoelastic fluids, or even active (e.g. bacterial) baths. 

\begin{acknowledgments}
SL acknowledges funding from the Deutsche Forschungsgemeinschaft (DFG, German Research Foundation) -- through the project 498288081. We thank Yue Liu for providing feedback on the final draft.
\end{acknowledgments}


\section{Methods}

\subsection{MD simulations}

We employed the LAMMPS for performing all out MD simulations~\cite{plimpton1995fast}. The interatomic force between atoms was accounted by a Lennard-Jones (LJ) potential function, 
\begin{equation}
\phi(r_{ij})=4\varepsilon[(\sigma/r_{ij} )^2- (\sigma/r_{ij} )^6 ],
\end{equation} where $r_{ij}$ is the interatomic distance between atom $i$ to atom $j$,  $\varepsilon$ is the depth of the potential well, and $\sigma$  the distance at which the particle--particle potential energy vanishes. The parameter values $\varepsilon$ and $\sigma$ of both Argon--Argon and Copper--Copper interactions are summarized in Table~\ref{tab:1}.
\begin{table}[ht]
\centering
\begin{tabular}[t]{lcc}
\hline
& $\varepsilon (\mathrm{eV})$ & $\sigma (\AA)$ \\
\hline
Ar – Ar&0.0104&3.405\\
Cu – Cu&0.4093&2.338\\
\hline
\end{tabular}
\caption{LJ interaction parameters \cite{zarringhalam2019effects}\label{tab:1}}
\end{table}%
For the interatomic forces between Argon and Copper atoms, we employed the Lorentz-Berthelot mixing rules~\cite{delhommelle2001inadequacy}, i.e.
\begin{eqnarray} 
\varepsilon_{12} = \sqrt{\varepsilon_1\: \varepsilon_2} , \hspace{1cm} \sigma_{12}=  (\sigma_2+ \sigma_1)/2 .
\end{eqnarray}

As shown in Fig.~\ref{fig:MD}, our setup consists of two containers of cold and hot (defined by blue and red colours respectively) Argon-Copper mixture, each with volume $11 \rm{nm}^3$ and containing $25280$ Ar fluid atoms with total mass $m _{\rm Ar}= 1.67\times 10^{-21} \rm{kg}$. The  nanoparticles are made each of  $186$ Cu  atoms  with nanoparticle mass $m= 1.96\rm{e}-23 \rm{kg}$ and radius $r = 1.44 \rm{nm}$. To prevent  interactions between the two containers and separate cold and hot baths, one-atom-thick Cu walls are placed between the containers, in the edges and in the middle  of the simulation box along the $X$ direction. Ar atoms of one container do not interact with the Ar atoms inside the other container.
Periodic boundary condition were applied in the $Y$ and $Z$ directions while the $X$ direction is constrained by the walls.
The simulations were carried out for $15-30 \rm{ns}$. Within this time the temperature of fluids at both hot and cold baths were controlled at $T_\mathrm{H}=120\rm{K}$ and $T_\mathrm{C}=100\rm{K}$ using  Nos\'e---Hoover thermostats (NVT)  with a coupling time-constant of $1\rm{ps}$. We have also checked that the reversed heat flow setup is not sensitive to the choice of Nos\'e---Hoover as it is also reproduced using a Langevin-based thermostat as well (data not shown).  A time step of $1 \rm{fs}$ was chosen for the MD simulations and the data was sampled every $1\rm{ps}$.

\subsection{Equations of motion}
The stochastic model that we use is given by two coupled underdamped Langevin equations for the  positions $x_\mathrm{C}$ and $x_\mathrm{H}$ of the center of mass of the nanoparticles along the $x$ axis
\begin{eqnarray} \label{eq:LeStart}
\hspace{-0.5cm}m  \ddot{x}_\mathrm{C}  &=& -  \gamma_\mathrm{C} \dot{x}_\mathrm{C} - \kappa x_\mathrm{C} + \kappa_\mathrm{C} (X_\mathrm{H} -x_\mathrm{C}-R ) + \xi_\mathrm{C}\\
\hspace{-0.5cm}m  \ddot{x}_\mathrm{H}&=& - \gamma_\mathrm{H} \dot{x}_\mathrm{H}   -\kappa(X_\mathrm{H}-R) -\kappa_\mathrm{H} (
x_\mathrm{H} - x_\mathrm{C} - R )+ \xi_\mathrm{H}.\nonumber\\
\end{eqnarray}
The parameters are defined in the Main Text below Eq.~\eqref{eq:Le}. 
Introducing the variables,
${x}_\mathrm{H}= {X}_\mathrm{H} - R$ and ${x}_\mathrm{C}= {X}_\mathrm{C}$, we can simplify \eqref{eq:LeStart} to
\begin{align}
m \ddot{X}_\mathrm{C}+\gamma_\mathrm{C} \dot{X}_\mathrm{C} &=  -(\kappa_\mathrm{C} + {\kappa} ) X_\mathrm{C} + \kappa_\mathrm{C} X_\mathrm{H} + \xi_\mathrm{C}
\\
m \ddot{X}_\mathrm{H}+\gamma_\mathrm{H} \dot{X}_\mathrm{H} &
=
\kappa_\mathrm{H}
X_\mathrm{C} -({\kappa}+\kappa_\mathrm{H}) X_\mathrm{H} + \xi_\mathrm{H},
\end{align}
as used throughout in the Main Text.
In all our MD simulations we set the stiffness of the  traps  to the value ${\kappa}=1\,\mathrm{eV/(nm)^2} = 0.16021\,\mathrm{J/m^2}=1.16 k_\mathrm{B}T_\mathrm{C}/\AA ^2$.
We further fix $\kappa_\mathrm{C}=10 \kappa $ and vary $\kappa_\mathrm{H}$.

\subsection{Estimation of the effective friction coefficient}
The noise terms as well as the friction forces appearing in the Langevin equations are not explicitly present in the MD simulations, but they emerge implicitly from the interactions between nanoparticle and surrounding bath particles. %
In our stochastic model, we assume Stokes law, in particular, instantaneous effective friction force that is linearly proportional on the instantaneous velocity, see Eq.~\eqref{eq:LeStart}. 
We obtain estimates for the values of the corresponding friction coefficients $\gamma_\mathrm{C,H}$ from MD simulations in the case of no coupling, $\kappa_\mathrm{C,H}=0$, via two distinct routes. 
First, we measure the velocity and positional autocorrelation functions in an equilibrium MD simulation, and fit the corresponding analytical expression taken from Ref.~\cite{wang1945theory}:
\begin{eqnarray}
\!\!\frac{\langle {X}_\mu (t){X}_\mu(t+\tau)\rangle}{\langle {X}_\mu^2 (t)\rangle}\!\! &=& \!\!\exp\left(-\frac{\gamma_\mu \tau}{2 m}\right) \left[ \cos( \omega_\mu \tau) + \frac{\gamma\sin(\omega_\mu \tau)}{2 m \omega_\mu}  \right]\nonumber\\
\!\!\frac{\langle {\dot{X}}_\mu (t){\dot{X}}_\mu(t+\tau)\rangle}{\langle {\dot{X}}_\mu^2 (t)\rangle}\!\! &=& \!\! \exp\left(-\frac{\gamma_\mu \tau}{2 m}\right) \left[ \cos( \omega_\mu \tau) -\frac{\gamma \sin(\omega_\mu \tau )}{2 m \omega_\mu} \right],\nonumber\\ \label{acfs}
\end{eqnarray}
with  $\omega_\mu^2 =(\kappa/m)-(\gamma_\mu/m)^2$, and we recall $\mu=\{\rm{C,H}\}$. 
Equations~\eqref{acfs} reproduce our numerical estimates for both the position and velocity autocorrelation functions in equilibrium conditions (see the SI for further details). Fitting~\eqref{acfs} to our numerical results by setting $m$, $\kappa$ to their input values in the simulations, we extract the estimates  $\gamma_\mathrm{C}\simeq \gamma_\mathrm{H}  \sim 3.5 \times 10^{-12}$ kg/s for the effective friction coefficient.
 We use this estimate for $\gamma$ and set $\gamma_\mathrm{C} = \gamma_\mathrm{H}$.
Second, we have performed a dragging experiment, pulling the nanoparticle through the bath and measure the resulting velocity (see Appendix for further details), which yields an estimate of $\gamma_\mathrm{C} \approx \gamma_\mathrm{H} \approx 3\times 10^{-12} \text{kg/s}$.
Both routes yield consistent results. We use $\gamma=\gamma_\mathrm{C} = \gamma_\mathrm{H} = 3.5\times 10^{-12} \text{kg/s}$ throughout the paper.

\subsection{Definition of heat and work in Stochastic Thermodynamics}
We address the thermodynamics of the nanoparticle system by applying the framework of stochastic thermodynamics~\cite{Sekimoto2010,Seifert2012} to our stochastic model. For convenience,  we first rewrite Eqs. (1,2) as a two-dimensional Langevin equation
\begin{equation}
    m\ddot{{\bf X}} = -\gamma \dot{{\bf X}} -\nabla E+ {\bf F}_{\rm nr} + \Xi
    \label{LE2}
\end{equation}
where ${\bf X}= ( X_\mathrm{C} , X_\mathrm{H} )^{\mathsf{T}}$, $\Xi=(\xi_\mathrm{C},\xi_\mathrm{H})^{\mathsf{T}}$, and $\nabla = (\partial_{X_\mathrm{C}},\partial_{X_\mathrm{H}})^{\mathsf{T}}$, with ${\mathsf{T}}$ denoting transposition. The term ${\bf F}_{\rm nr}=(\kappa_\mathrm{C} (X_\mathrm{H}-R), \kappa_\mathrm{H} X_\mathrm{C})^{\mathsf{T}}$ is the nonreciprocal force acting on the system, which is non-conservative. On the other hand,  the energy $E$ of the nanoparticles 
\begin{equation}
    E = \frac{m}{2}(\dot{X}_\mathrm{C}^2 + \dot{X}_\mathrm{H}^2) + \left(\frac{\kappa+\kappa_\mathrm{C}}{2}\right) X_\mathrm{C}^2 + \left(\frac{\kappa+\kappa_\mathrm{H}}{2}\right)   (X_\mathrm{H}-R)^2 
\end{equation}
is a function  of the instantaneous values of the particles' positions and velocities, and thus a quantity that fluctuates during a simulation,  i.e. a stochastic process. 
  Following Sekimoto~\cite{Sekimoto2010}, the energy change in a small time interval $[t,t+\rm{d}t]$ can be written as
\begin{equation}\label{def:u}
    \mathrm{d}E = \nabla E({\bf X}) \circ \mathrm{d}{\bf X} + m\ddot{{\bf X}}\circ \mathrm{d}{{\bf X}},
\end{equation}
where $\circ$ denotes the Stratonovich product. Here, $\mathrm{d}{\bf X}= ( \mathrm{d}X_\mathrm{C} , \mathrm{d}X_\mathrm{H} )^T  $ is the stochastic increment of the nanoparticles' positions which follows from the Langevin dynamics~\eqref{LE2}. In $[t,t+\rm{d}t]$ the stochastic work done on the system is given by the non-conservative force times the displacement of the nanoparticles, which we can split as
\begin{equation}\label{def:w}
    \mathrm{d}W = {\bf F_{\rm nr}}\circ  \mathrm{d}{{\bf X}} = \underbrace{\kappa_\mathrm{C} (X_\mathrm{H}-R) \circ  \mathrm{d}X_\mathrm{C}}_{=\displaystyle \mathrm{d}W_\mathrm{C}} +  \underbrace{\kappa_\mathrm{H} X_\mathrm{C}\circ  \mathrm{d}X_\mathrm{H}}_{=\displaystyle \mathrm{d}W_\mathrm{H}},
\end{equation}
where $\mathrm d W_\mathrm{C}$ ($\mathrm d
W_\mathrm{H}$) is the work done on the cold (hot) nanoparticle. 
Dividing the stochastic work $\mathrm{d}W$ applied to the particle along a trajectory of length by the trajectory length $\mathrm{d}t$ yields the power $\dot{W}= \mathrm{d}{W}/\mathrm{d}t$.

Similarly, the stochastic heat dissipated in the same time interval 
\begin{equation}
    \mathrm{d} Q = \mathrm{d}E - \mathrm{d}W =  \underbrace{(\gamma_\mu \dot{X}_\mathrm{C} - \xi_\mathrm{C} )\circ \mathrm{d}{X}_\mathrm{C}}_{=\displaystyle \mathrm{d}Q_\mathrm{C}} + \underbrace{(\gamma_\mu \dot{X}_\mathrm{H} - \xi_\mathrm{H} )\circ \mathrm{d}{X}_\mathrm{H}}_{=\displaystyle \mathrm{d}Q_\mathrm{H}} ,
    \label{def:heat}
\end{equation}
which ensures that the first law is satisfied for every single trajectory traced by the system. 
Note that we use the thermodynamic sign convention: $W>0$ ($W<0$) when work is exerted (extracted) from the system and $Q>0$ ($Q<0$) when heat is dissipated from (absorbed by) the system to (from) its environment. 
Dividing the stochastic heat dissipated along a trajectory of length by the trajectory length $\rm{d} t$ yields the stochastic heat dissipation rate $\dot{Q}= {\rm{d}} {Q}/\mathrm{d}t$.

\subsection{Power fluctuations}
The power fluctuations of each nanoparticle predicted by our stochastic model are given in   Eq. \eqref{eq:Pw} as closed-form expressions for $P(\dot W_j)$ with $j\in \{ \text{C,H} \}$. The distributions $P(\dot W_j)$  explicitly depend on $\alpha_j,\beta_j,\zeta_j$, which are in turn functions of the covariances of the positions and velocities of the $j$ nanoparticle  and the $l\neq j$ nanoparticle. Specifically, %
\begin{align}\label{def:alpha}
\alpha_j &= \frac{1}{2 \langle X_l^2 \rangle (1-\psi_j^2)},
~~
\beta_j = \frac{\psi_j}{\sqrt{\langle  X_l^2 \rangle\langle \dot X_j^2 \rangle}(1-\psi^2_j)},
\end{align}
and
\begin{align}
~~\zeta_j = \frac{1}{2 \langle \dot X_j^2 \rangle (1-\psi_j^2)}, ~~
\psi_j=\frac{\langle X_l \dot{X}_j \rangle }{\sqrt{\langle  X_l^2 \rangle\langle \dot X_j^2 \rangle}},
\end{align}
with
\begin{align}
\langle  {X}_l^2 \rangle
&=
k_\mathrm{B}
\frac{
m T_l [\kappa_j^2 (\kappa_j+\kappa_l) + \kappa (\kappa_j^2 + \kappa_l^2)]}
{\kappa(\kappa + \kappa_j + \kappa_l)[m \left[(\kappa_j + \kappa_l)^2 +2 (2\kappa + \kappa_j + \kappa_l)\right]]}
\nonumber \\
&+k_\mathrm{B}
\frac{ 2\gamma^2 [\kappa_l^2 T_j + T_l(\kappa+\kappa_j)(2\kappa+\kappa_j) +\kappa \kappa_l T_l]}
{\kappa(\kappa + \kappa_j + \kappa_l)[m \left[(\kappa_j + \kappa_l)^2 +2 (2\kappa + \kappa_j + \kappa_l)\right]]}
\nonumber \\
&+
k_\mathrm{B}
\frac{
m T_j \kappa_l^2 [2 \kappa + \kappa_j + \kappa_l]
}
{\kappa(\kappa + \kappa_j + \kappa_l)[m \left[(\kappa_j + \kappa_l)^2 +2 (2\kappa + \kappa_j + \kappa_l)\right]]},
\end{align}
\begin{align}
\langle  \dot{X}_j^2   \rangle
=
k_\mathrm{B}\frac{ \kappa_j (\kappa_j T_l-\kappa_l T_j)}{(m/2) (\kappa_j + \kappa_l)^2+ \gamma ^2 (2\kappa+\kappa_j+\kappa_l)}+\frac{k_\mathrm{B} T_j}{m},
\end{align}
and 
\begin{align}
\langle  {X}_l{\dot{X}}_j \rangle
=
k_\mathrm{B}
\frac{
2\gamma(\kappa_j T_l - \kappa_l T_j )}
{m (\kappa_j+\kappa_l)^2+ 2(2\kappa+ \kappa_j \kappa_l) \gamma ^2}.
\end{align}
From the power distributions $P(\dot W_j)$, we can further deduce analytical expressions for the variance of the power:
\begin{align}\label{eq:PwVar}
    \text{Var}(\dot W_j)& =
    \frac{1}{\mathcal{Z}_j}\!\left\{ \frac{2\kappa_j^3}{(2\sqrt{\alpha_j \zeta_j}+\beta_j)^3} +
    \frac{2\kappa_j^3}{(2\sqrt{\alpha_j \zeta_j}-\beta_j)^3}
    \right.\nonumber \\&- \left.
    \left[ \frac{\kappa_j^2}{(2\sqrt{\alpha_j \zeta_j}+\beta_j)^2} +
    \frac{2\kappa_j^2}{(2\sqrt{\alpha_j \zeta_j}-\beta_j)^2}\right]^2 \right\} .
\end{align}
See the Supplemental Material for further details and the explicit mathematical derivations.

\bibliography{References.bib}


\onecolumngrid
\appendix
\setcounter{secnumdepth}{2}
\newpage
\begin{center}
	\Large{\textbf{Supplemental Material}}
\end{center}

\section{Estimation of friction coefficients}

\subsection{Dragging experiment in MD simulations}

\begin{figure}[h!]
	\centering
	\includegraphics[width=0.75\textwidth]{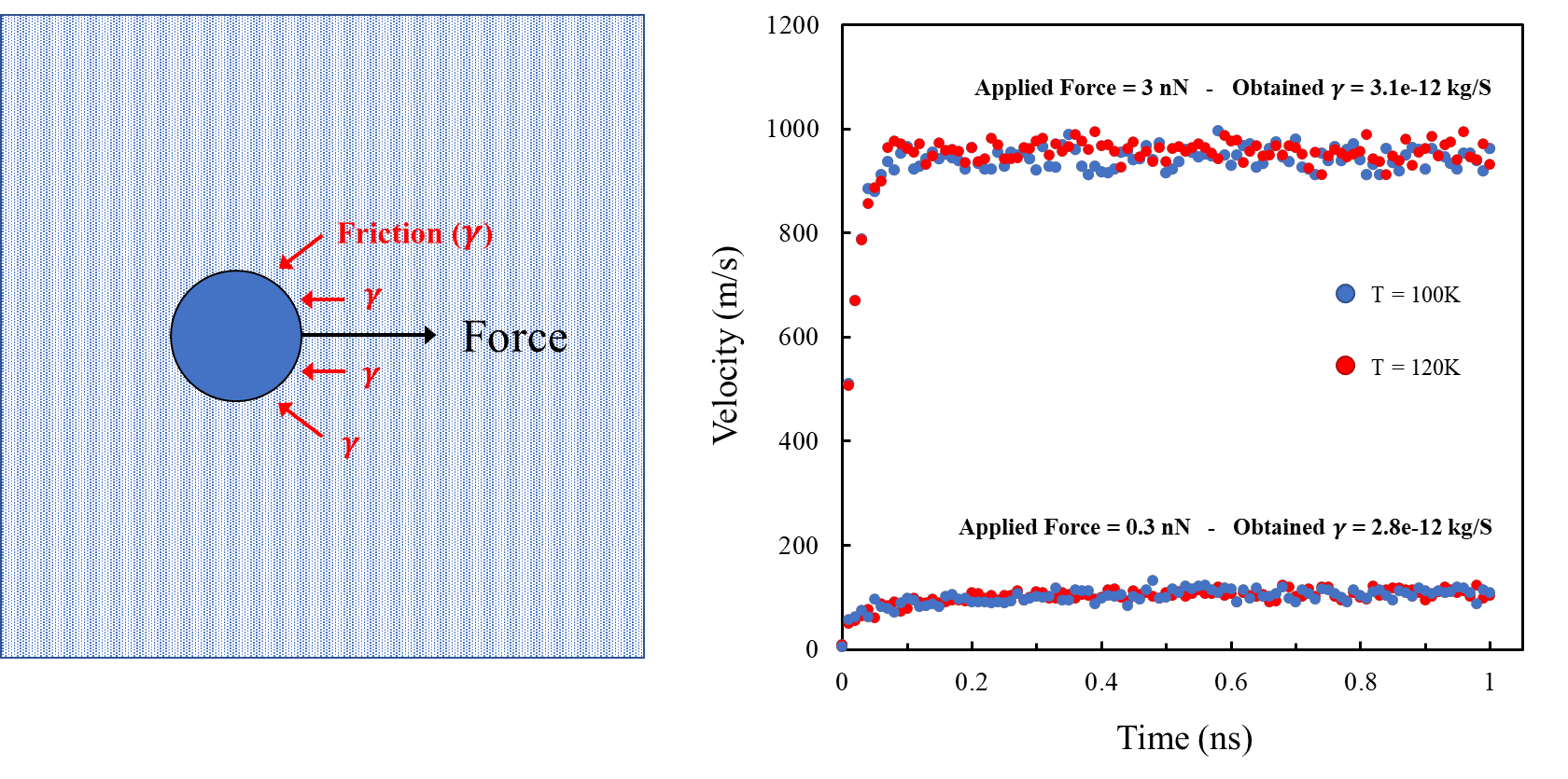}
	\caption{Left: Sketch of the dragging  MD simulation used to estimate the friction coefficient (see text for details).  Right: instantaneous velocity of the center of mass of the nanoparticle Assuming Stokes' law for the friction force $F_\mathrm{friction}=-\gamma v$, an estimate for the friction coefficient of $\gamma \sim 3\times 10^{-12} $kg/s.}\label{fig:dragging}
\end{figure}

To estimate the friction coefficient, a dragging MD simulation is performed assuming the Stokes' law for the friction force ($F_\mathrm{friction}=-\gamma v$). In this regard, a constant force along the $x$ axis of magnitude $F = 0.3 $nN is applied to a spherical copper nanoparticle with radius $r = 1.44$nm immersed in an  Ar fluid bath with fixed temperature  $100$K (first simulation) and $120$K (second simulation) and periodic boundary conditions. As shown in Fig.~\ref{fig:dragging}, the velocity  of the Cu nanoparticle reached a steady value (terminal velocity) after a transient time of $\sim 0.3$ns and remained constant for the rest of the simulation. Using the Stokes' law, the friction coefficient is estimated from the ratio between the magnitude of the external force and the terminal velocity $\gamma=~2.8\times 10^{-12} $kg/s obtained as the mean of the simulations done at the two  temperature values. The simulation is repeated with an external force ten times  larger ($F = 3$nN), yielding a very similar estimate of  the friction coefficient  $\gamma=~3.1\times 10^{-12}$kg/s.

\newpage

\subsection{Equilibrium autocorrelation functions: MD simulations and  Langevin theory}

\begin{figure}[H]
	\centering
	{\large{\textsf{(a)}}}~~~~~~~~~~~~~~~~~~~~~~~~~~~~~~~~~~~~~~~~~~~~~~~~~~~~~~~~~~~~~{\large{\textsf{(b)}}}\\
	\includegraphics[width=0.35\textwidth]{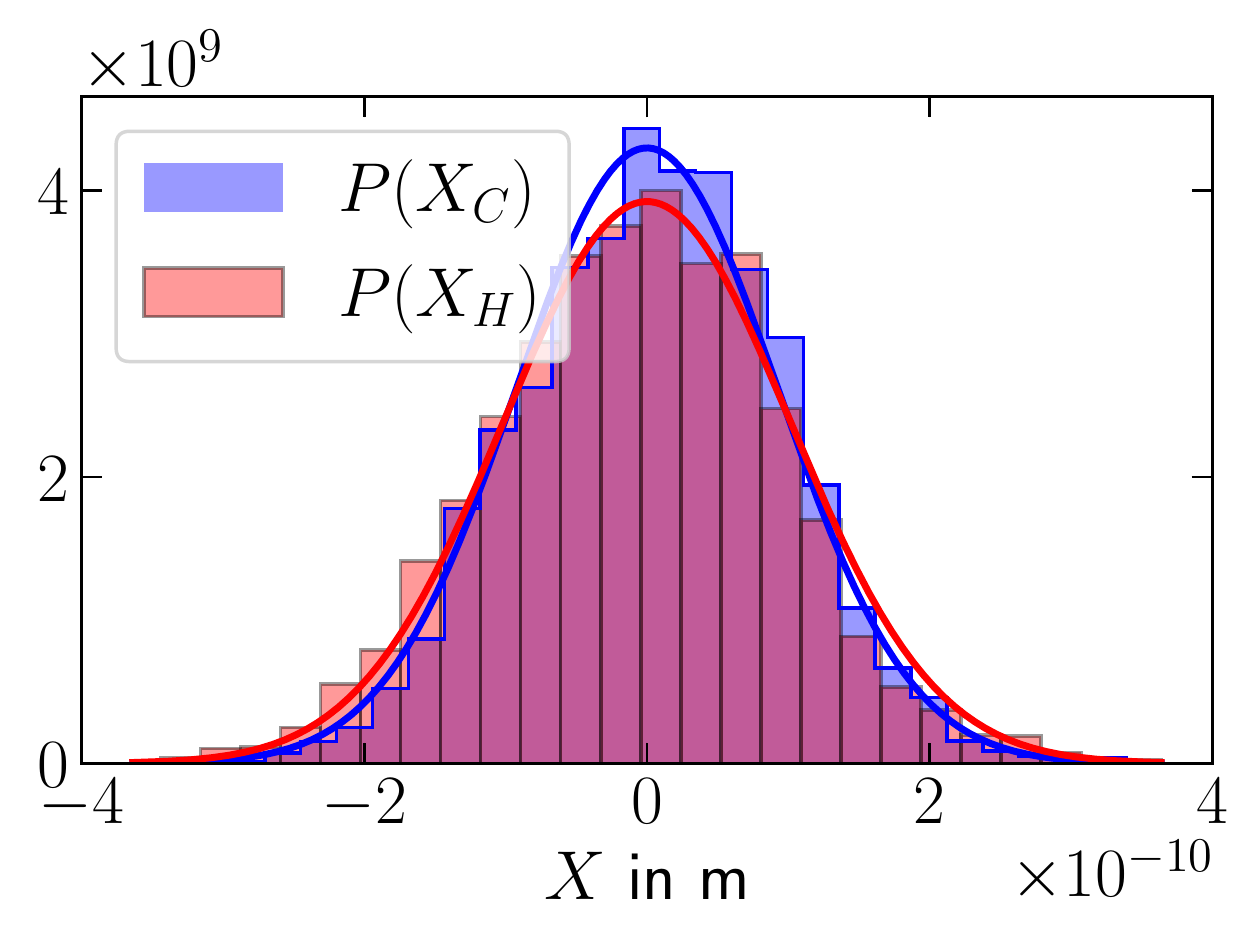}
	\includegraphics[width=0.35\textwidth]{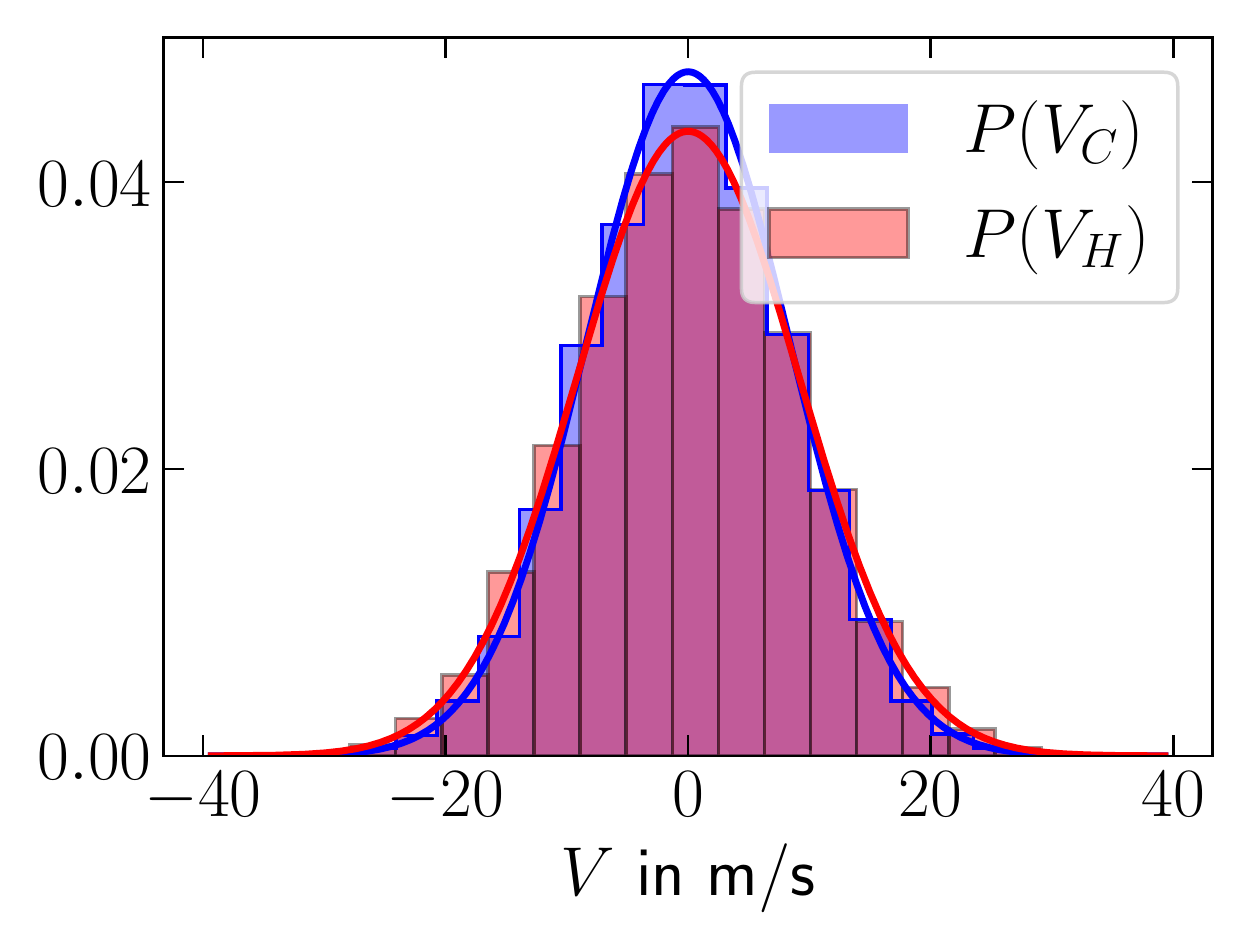}
	\\
	{\large{\textsf{(c)}}}~~~~~~~~~~~~~~~~~~~~~~~~~~~~~~~~~~~~~~~~~~~~~~~~~~~~~~~~~{\large{\textsf{(d)}}}~~~~~~~~~~~~~~~~~~~~~~~~~~~~~~~~~~~~~~~~~~~~{\large{\textsf{(e)}}}\\
	\includegraphics[width=0.3\textwidth]{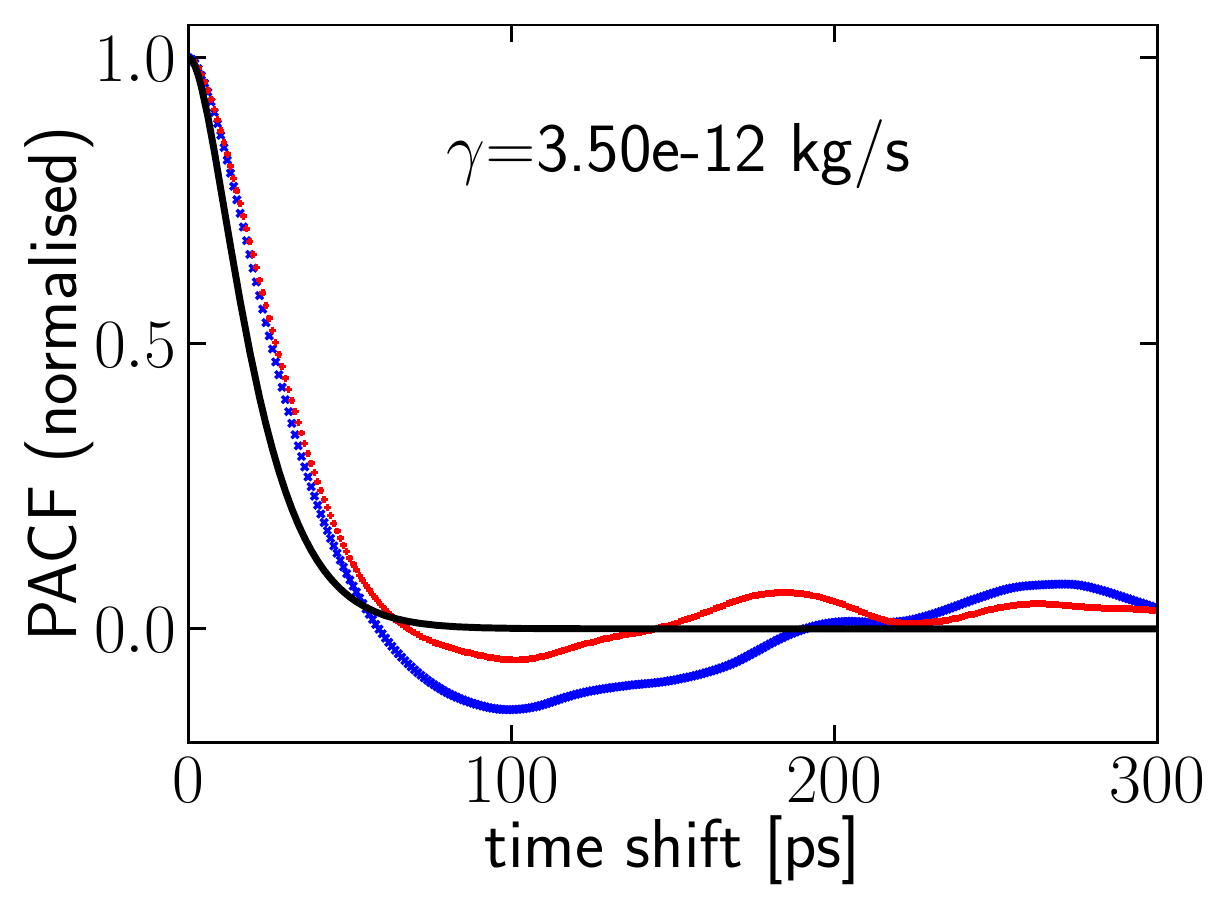}
	\includegraphics[width=0.3\textwidth]{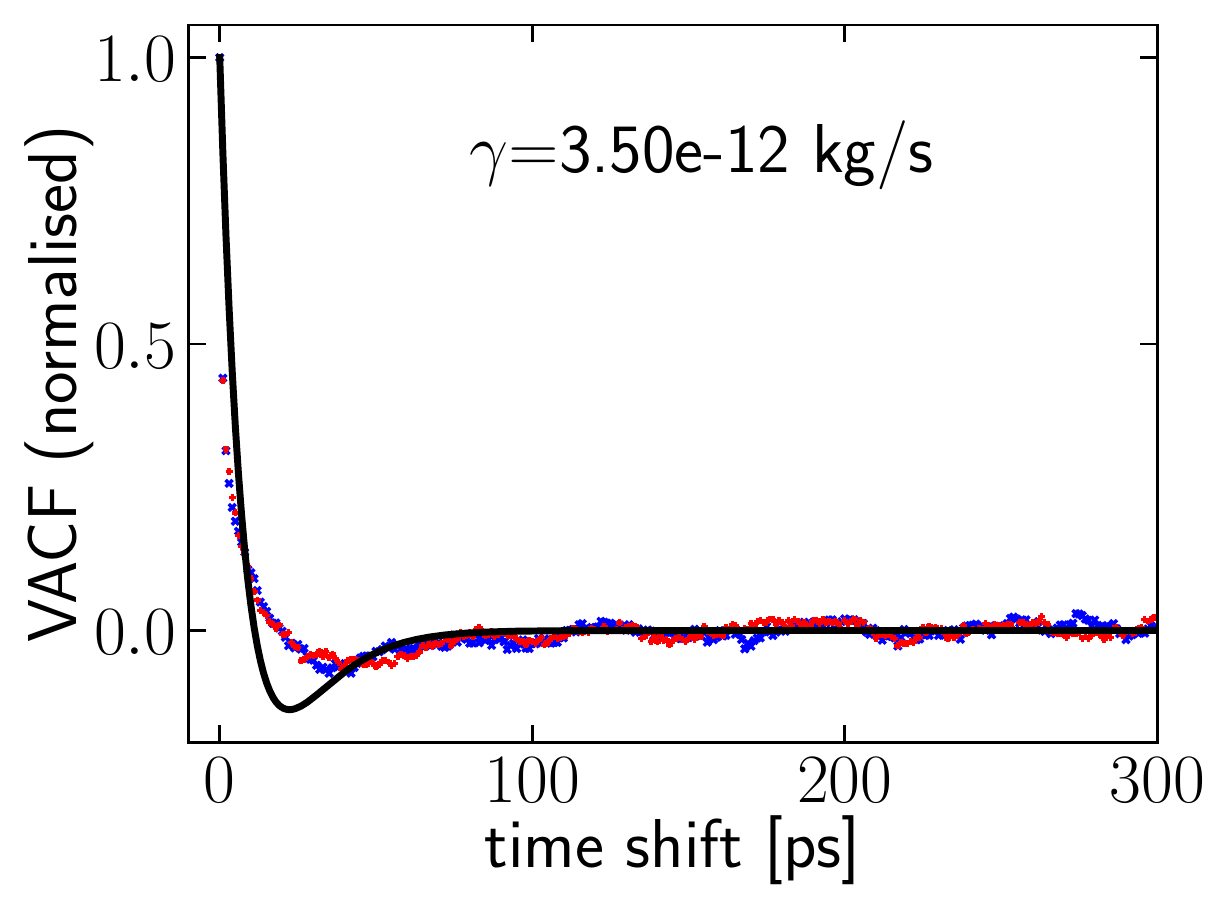}
	\includegraphics[width=0.3\textwidth]{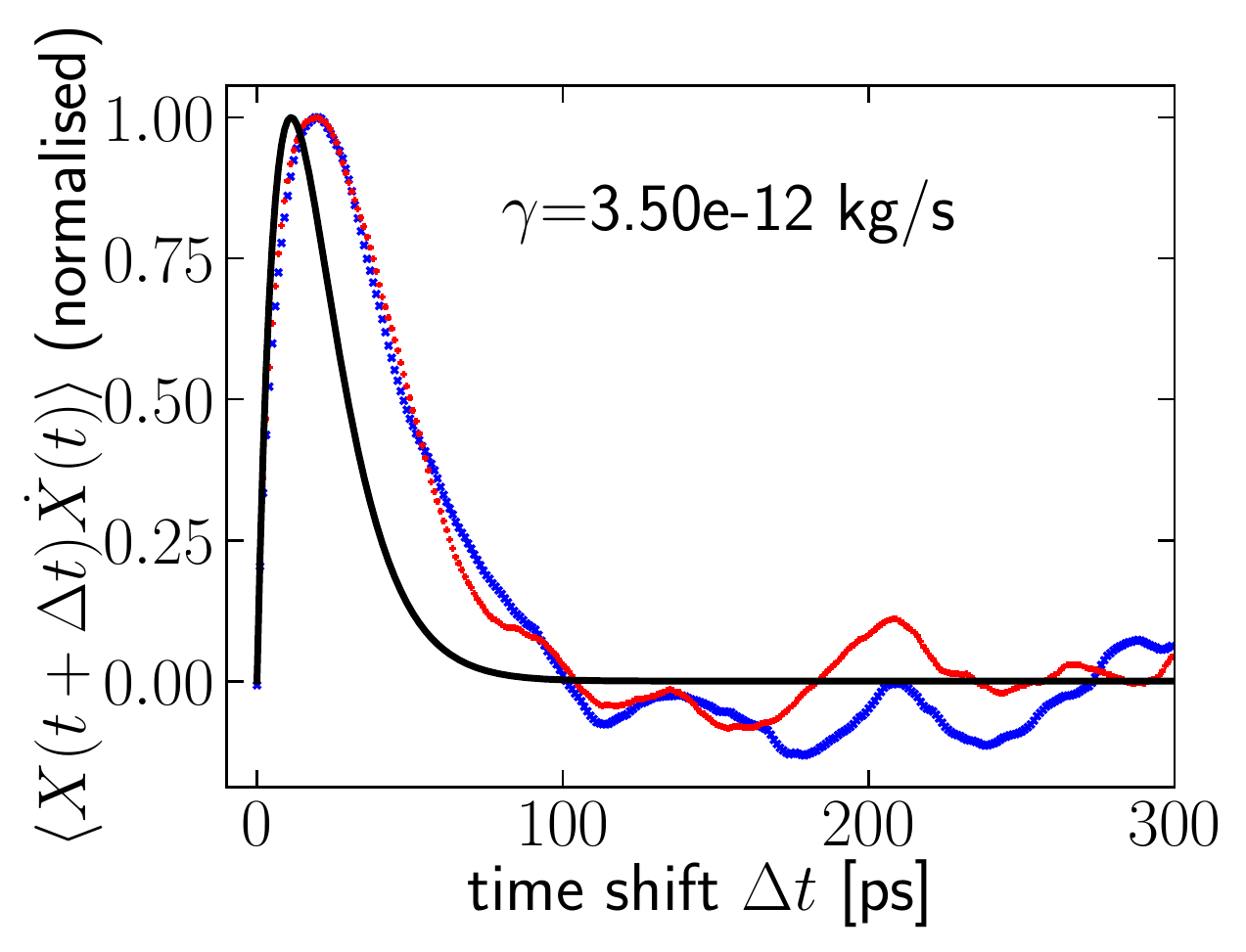}
	\caption{
		(a,b) Positional and velocity fluctuations of nanoparticles in the uncoupled equilibrium case ($\kappa_\mathrm{H} = \kappa_\mathrm{C} =0$). The bars are obtained from MD simulations and the lines show the theoretical predictions, i.e., zero-mean Gaussian functions with variance $k_\mathrm{B}T/a$, $k_\mathrm{B}T/m$, respectively, see Eq.~\eqref{eq:b3}.
		(c-e) Position autocorrelation function (PACF, c) and velocity autocorrelation function (VACF, d) and position-velocity cross correlation function (e) obtained from equilibrium MD simulations (with $\kappa_\mathrm{C}=\kappa_\mathrm{H}=0$) extracted from the center-of-mass motion of the hot (red line) and cold nanoparticle (blue line). The black lines in (c-e) are fits to the analytical expressions for the underdamped Langevin model: Eq.~\eqref{eq:PACF} (c),  Eq.~\eqref{eq:VACF} (d) and  Eq.~\eqref{eq:VC-Corr} (e)  to the theoretical curves \eqref{eq:VC-Corr}.
		The insets show the value of the effective friction extracted from the fits with only one free parameter, all of them yielding the estimate $\gamma \sim 3.5 \times 10^{-12}$kg/s.
	}\label{fig:PACF}
\end{figure}

Next, we describe how we extracted the effective friction coefficients from fits of the autocorrelation functions obtained from the MD simulations to analytical expressions derived below for our underdamped Langevin model.
In thermal equilibrium, the normalised autocorrelation functions of the particle position and velocity are known. For $\kappa_\mathrm{C}=\kappa_\mathrm{H}=0$, they read \cite{wang1945theory}
\begin{equation}\label{eq:PACF}
	\text{PACF}(\tau) = \exp\left(-\frac{\gamma \tau}{2 m}\right) \left[ \cos( \omega_1 \tau) + \frac{\gamma}{2 m \omega_1} \sin(\omega_1 \tau ) \right],
\end{equation}
and
\begin{align}\label{eq:VACF}
	\text{VACF}(\tau) = \exp\left(-\frac{\gamma \tau}{2 m}\right) \left[ \cos( \omega_1 \tau) - \frac{\gamma}{2 m \omega_1} \sin(\omega_1 \tau ) \right],
\end{align}
with $\omega_1 = \sqrt{\frac{\kappa}{m}-\frac{\gamma^2}{m^2}}$. Further, the position--velocity cross correlation is given by
\begin{equation}
	\label{eq:VC-Corr}
	\langle X(t)\dot{X}(t+\tau) \rangle = \exp\left(-\frac{\gamma \tau}{2 m}\right) \left [ \frac{\gamma\sin(\omega_1 \tau )}{2 m \omega_1} \right ].
\end{equation}
Figure \ref{fig:PACF} displays  MD simulation results from equilibrium simulations together with the fits to the theoretical  predictions given by Eqs.(\ref{eq:PACF}-\ref{eq:VC-Corr}).

\section{Analytical calculation of heat and entropy production rates \label{sec:heatCalculation}}
\subsection{Heat flow rates}
\label{sec:heatflowrates}
Here we calculate the ensemble-averaged heat flow rates $   \langle \dot{Q}_j   \rangle$, with $j\in\{{\mathrm{C},\mathrm{H}}\}$. To this end, we start from  Sekimoto's definition~\cite{Sekimoto2010} of the stochastic heat dissipated to the $j$ bath in $[t,t+\text{d}t]$, i.e.  $\mathrm{d}Q_j = (\gamma\dot{X}_j - \xi_j)\circ \mathrm{d}X_j$, 
and consider the steady-state  average
\begin{align}
	\langle \dot{Q}_j \rangle=  \left \langle \frac{\mathrm{d}Q_j}{\mathrm{d}t} \right \rangle =\gamma \langle \dot{X}_j^2 \rangle - \left \langle 
	\xi_j\dot{X}_j \right \rangle.
\end{align}
Inserting $\langle  \xi_j\dot{X}_j \rangle = \gamma k_\mathrm{B} T_j /m$, which follows from the Langevin equation \cite{Loos2019}, one finds 
\begin{align}\label{eq:heatvelocity}
	\langle\dot{Q}_j\rangle = \gamma \left[ \langle \dot{X}_j^2   \rangle - \frac{k_\mathrm{B}T_j}{m} \right] .
\end{align}
Thus, the heat flow between each nanoparticle and its bath is directly given by the variance of its velocity fluctuations. We note that this formula can also be recast into the form $\langle \dot{Q}_j \rangle = \frac{k_\mathrm{B}\gamma}{m} \left[ 
T_j^\mathrm{eff}
- T_j \right]$, i.e., the heat flow is proportional to the difference between ``effective temperature'' $T_j^\mathrm{eff}=m\langle \dot{X}_j^2   \rangle/k_\mathrm{B}$ and actual temperature $T_j$.

The stationary joint probability density of the position and velocity of the two particles described by the linear Langevin equation~\eqref{eq:Le}
is given by the four-variate normal distribution
\begin{equation}\label{eq:b3}
	\rho_4 (x_\mathrm{C},x_\mathrm{H},v_\mathrm{C},v_\mathrm{H}) = \frac{\exp\left[-\frac{1}{2}(\mathbf{r}-\langle\mathbf{r}\rangle)^{\mathsf{T}}\mathbf{C}^{-1}(\mathbf{r}-\langle\mathbf{r}\rangle)\right]}{\sqrt{(2\pi)^4\text{det}\mathbf{C}}},
\end{equation}
where $\mathbf{r}=(x_\mathrm{C},x_\mathrm{H},v_\mathrm{C},v_\mathrm{H})^{\mathsf{T}}$ is a column vector with $\mathsf{T}$ denoting matrix transposition, $\langle\mathbf{r}\rangle$ its stationary average, and $\mathbf{C}$ the correlation matrix which is given by
\begin{equation}
	\mathbf{C}=\begin{pmatrix}
		\langle  {X}_\mathrm{C}^2   \rangle &  \langle  {X}_\mathrm{C} {X}_\mathrm{H}   \rangle 
		& \langle  {X}_\mathrm{C}\dot{X}_\mathrm{C}   \rangle &  \langle {X}_\mathrm{C}\dot{X}_\mathrm{H}   \rangle
		\\
		\langle  {X}_\mathrm{C} {X}_\mathrm{H}   \rangle &  \langle {X}_\mathrm{H}^2   \rangle 
		& \langle  {X}_\mathrm{H}\dot{X}_\mathrm{C}   \rangle &  \langle {X}_\mathrm{H}\dot{X}_\mathrm{H}   \rangle 
		\\
		\langle  \dot{X}_\mathrm{C}{X}_\mathrm{C}   \rangle &  \langle \dot{X}_\mathrm{C} {X}_\mathrm{H}   \rangle 
		& \langle  \dot{X}_\mathrm{C}^2   \rangle & \langle \dot{X}_\mathrm{C}\dot{X}_\mathrm{H}   \rangle
		\\
		\langle  \dot{X}_\mathrm{H}{X}_\mathrm{C}   \rangle &  \langle \dot{X}_\mathrm{H} {X}_\mathrm{H}   \rangle 
		&  \langle \dot{X}_\mathrm{C}\dot{X}_\mathrm{H}   \rangle &\langle  \dot{X}_\mathrm{H}^2   \rangle
	\end{pmatrix}. 
\end{equation}

To calculate the variance of the velocity fluctuations, we develop further  the formalism introduced in Refs.~\cite{kwon2005structure,bae2021inertial,kwon2011nonequilibrium} to calculate correlation matrices. 
To this end, we first recast the Langevin equations into the matrix equations
\begin{equation}\label{eq:LE-Matrix}
	\dot{\mathbf{r}}
	= -\mathbf{B}\,\mathbf{r} + \mathbf{\Xi} 
	=
	-\begin{pmatrix}
		\bf{0}_2 & -{\bf{I}_2}\\
		-\mathbf{A}/m & \mathbf{\Gamma}/m
	\end{pmatrix} 
	\,\mathbf{r} + \mathbf{\Xi} 
	,~~~\mathbf{\Gamma}=\begin{pmatrix}
		\gamma & 0\\
		0 & \gamma
	\end{pmatrix}
	,~~~\mathbf{A}=\begin{pmatrix}
		a_0 & k_0\\
		k_1 & a_1
	\end{pmatrix}
	,
\end{equation}
with the state vector $\mathbf{r}=(X_\mathrm{C},X_\mathrm{H},\dot{X}_\mathrm{C},\dot{X}_\mathrm{H})^{\mathsf{T}}$ and the noise vector $\mathbf{\Xi}=(0,0,\xi_\mathrm{C},\xi_\mathrm{H})^{\mathsf{T}}$. The $4 \times 4$ matrix 
$\mathbf{B}$
contains the $2\times 2$ zero and identity matrices, $\bf{0}_2$ and ${\bf{I}_2}$, the friction (diagonal) matrix
$\mathbf{\Gamma}$
and the 
coupling matrix $\mathbf{A}$.
For the sake of generality, we will here consider the most general case of coupling here (four independent coupling  entries of the coupling matrix), and below specialize our results to 
$\mathbf{A}=\begin{pmatrix}
	-(\kappa+\kappa_\mathrm{C}) & \kappa_\mathrm{C}\\
	\kappa_\mathrm{H} & -(\kappa+\kappa_\mathrm{H})
\end{pmatrix}.$
%
Further, we define the diffusion matrices 
$\mathbf{D}_2=\frac{k_\mathrm{B}\gamma}{m^2}\begin{pmatrix}
	T_\mathrm{C} &0\\
	0 & T_\mathrm{H}
\end{pmatrix}$ and $\mathbf{D}_4= \begin{pmatrix}
	\bf{0}_2  & \bf{0}_2 \\
	\bf{0}_2  & \mathbf{D}_2
\end{pmatrix}. $
From these ingredients, we can determine the correlation matrix
$
\mathbf{C}$ from the following expressions:
\begin{equation}
	\mathbf{C}=\mathbf{B}^{-1} (\mathbf{D}_4 + \mathbf{Q} )
\end{equation}
where $\mathbf{Q}$ is an anti-symmetric $4\times4$ matrix that is uniquely determined by
\begin{equation}
	\mathbf{B}\mathbf{Q}+\mathbf{Q}\mathbf{B}^{\mathsf{T}}=\mathbf{B}\mathbf{D}_4 - \mathbf{D}_4\mathbf{B}^{\mathsf{T}}.
\end{equation}
From this expression, and the fact that in stationary state $\langle \dot{X}_\mathrm{C}{X}_\mathrm{C} \rangle= \langle \dot{X}_\mathrm{H}{X}_\mathrm{H} \rangle=0 $, we obtain
\begin{subequations}
	\begin{align}
		\mathbf{C}=\begin{pmatrix}
			\langle  {X}_\mathrm{C}^2   \rangle &  \langle  {X}_\mathrm{C} {X}_\mathrm{H}   \rangle 
			& 0 &  \langle {X}_\mathrm{C}\dot{X}_\mathrm{H}   \rangle
			\\
			\langle  {X}_\mathrm{C} {X}_\mathrm{H}   \rangle &  \langle {X}_\mathrm{H}^2   \rangle 
			& \langle  {X}_\mathrm{H}\dot{X}_\mathrm{C}   \rangle & 0
			\\
			0 &  \langle \dot{X}_\mathrm{C} {X}_\mathrm{H}   \rangle 
			& \langle  \dot{X}_\mathrm{C}^2   \rangle & \langle \dot{X}_\mathrm{C}\dot{X}_\mathrm{H}   \rangle
			\\
			\langle  \dot{X}_\mathrm{H}{X}_\mathrm{C}   \rangle &  0
			&  \langle \dot{X}_\mathrm{C}\dot{X}_\mathrm{H}   \rangle &\langle  \dot{X}_\mathrm{H}^2   \rangle
		\end{pmatrix},
	\end{align} 
	with nonzero elements given by  
	\begin{align}
		\langle  \dot{X}_\mathrm{C}^2   \rangle
		=k_\mathrm{B}
		\frac{2 k_0 (k_0 T_\mathrm{H}-k_1 T_\mathrm{C})}{m \left[(a_0-a_1)^2+4 k_0 k_1\right]-2 \gamma ^2 (a_0+a_1)}+\frac{k_\mathrm{B} T_\mathrm{C}}{m},
	\end{align}
	\begin{align}
		\langle  \dot{X}_\mathrm{H}^2   \rangle
		=k_\mathrm{B}
		\frac{2 k_1 (k_1 T_\mathrm{C}-k_0 T_\mathrm{H})}
		{m \left[(a_0-a_1)^2+4 k_0 k_1\right]-2 \gamma ^2 (a_0+a_1)}+\frac{k_\mathrm{B} T_\mathrm{H}}{m},
	\end{align}
	\begin{align}
		\langle  \dot{X}_\mathrm{C}\dot{X}_\mathrm{H}  \rangle
		=
		\langle  \dot{X}_\mathrm{H}\dot{X}_\mathrm{C}  \rangle
		=
		k_\mathrm{B}
		\frac{(a_0-a_1) (k_1 T_\mathrm{C}-k_0 T_\mathrm{H})}{m \left[(a_0-a_1)^2+4 k_0 k_1\right]-2 \gamma ^2 (a_0+a_1)},
	\end{align}
	\begin{align}
		\langle  {X}_\mathrm{C}^2 \rangle
		=
		k_\mathrm{B}
		\frac{
			m T_\mathrm{C} [-(a_0-a_1)^2 a_1 + (a_0 - 3a_1)k_0k_1]
			-k_0^2 m T_\mathrm{H} (a_0+a_1) + 2\gamma^2 [a_1(a_0+a_1)T_\mathrm{C} - k_0 k_1 T_\mathrm{C} + k_0^2 T_\mathrm{H}]}
		{(a_0 a_1-k_0 k_1)[m \left[(a_0-a_1)^2+4 k_0 k_1\right]-2 \gamma ^2 (a_0+a_1)]},
	\end{align}
	\begin{align}
		\langle  {X}_\mathrm{C}{X}_\mathrm{H} \rangle
		=
		k_\mathrm{B}
		\frac{
			m T_\mathrm{C} k_1 [-a_0a_1+a_1^2+2k_0k_1]
			+k_0 m T_\mathrm{H} (a_0^2 - a_0 a_1 +2k_0k_1) + 2\gamma^2 (a_1 k_1 T_\mathrm{C} - a_0 k_0 T_\mathrm{H} )}
		{(a_0 a_1-k_0 k_1)[m \left[(a_0-a_1)^2+4 k_0 k_1\right]-2 \gamma ^2 (a_0+a_1)]},
	\end{align}
	\begin{align}
		\langle  {X}_\mathrm{C}{\dot{X}}_\mathrm{H} \rangle
		=
		k_\mathrm{B}
		\frac{
			2\gamma(k_1 T_\mathrm{C} - k_0 T_\mathrm{H} )}
		{m \left[(a_0-a_1)^2+4 k_0 k_1\right]-2 \gamma ^2 (a_0+a_1)}.
	\end{align}
\end{subequations}
This generalizes the results given in \cite{bae2021inertial} to the case of two independent trap stiffness and coupling strengths.
For the case considered here, 
$a_0 = -(\kappa + \kappa_\mathrm{C})$, $k_0 =\kappa_\mathrm{C}$, $a_1 = -(\kappa + \kappa_\mathrm{H})$, $k_1 =\kappa_\mathrm{H}$,
\begin{align}\label{eq:vc2}
	\langle  \dot{X}_\mathrm{C}^2   \rangle
	=
	k_\mathrm{B}\frac{ \kappa_\mathrm{C} (\kappa_\mathrm{C} T_\mathrm{H}-\kappa_\mathrm{H} T_\mathrm{C})}{(m/2) (\kappa_\mathrm{C} + \kappa_\mathrm{H})^2+ \gamma ^2 (2\kappa+\kappa_\mathrm{C}+\kappa_\mathrm{H})}+\frac{k_\mathrm{B} T_\mathrm{C}}{m}.
\end{align}

Hence, we find substituting~\eqref{eq:vc2} in Eq.~\eqref{eq:heatvelocity} 
\begin{equation}
	\langle \dot{Q}_\mathrm{C} \rangle = 
	k_\mathrm{B}\frac{ \kappa_\mathrm{C} (\kappa_\mathrm{C} T_\mathrm{H}-\kappa_\mathrm{H} T_\mathrm{C})}{(1/2)(m/\gamma) (\kappa_\mathrm{C} + \kappa_\mathrm{H})^2+ \gamma  (2\kappa+\kappa_\mathrm{C}+\kappa_\mathrm{H})} ,
\end{equation}
which coincides with Eq.~\eqref{eq:Qc} in the Main Text. The expression~\eqref{eq:Qh}  in the Main Text for the rate of heat dissipation to the hot bath can be found following analogous steps.

\subsection{Entropy production rate}
The corresponding total entropy production rate, is defined by~\cite{Seifert2012}
\begin{align}\label{eq:Stotdef}
	\langle 
	\dot{S}_\mathrm{tot}\rangle= 
	\langle \dot{Q}_\mathrm{H}/T_\mathrm{H}\rangle  +\langle \dot{Q}_\mathrm{C}/T_\mathrm{C}\rangle +\langle \dot{S}_\mathrm{sh}\rangle .
\end{align}
The last term denotes the rate of change of the Shannon entropy $\langle S_\mathrm{sh} \rangle =k_\mathrm{B}\langle -\ln \rho_4(X_\mathrm{C},X_\mathrm{H},\dot{X}_\mathrm{C},\dot{X}_\mathrm{H}) \rangle$ of the joint probability density function. This term is constant in the steady state, thus, $\langle \dot{S}_\mathrm{sh} \rangle=0$. In the present case, using \eqref{eq:Qc} and \eqref{eq:Qh}  and $\langle \dot{S}_\mathrm{sh} \rangle=0$ and Eq.~\eqref{eq:Stotdef}, we find
\begin{align}
	\langle \dot{S}_\mathrm{tot} \rangle
	& =\frac{k_\mathrm{B}}{T_\mathrm{C}T_\mathrm{H}}
	\frac{( \kappa_\mathrm{C}T_\mathrm{H}  - \kappa_\mathrm{H} T_\mathrm{C})^2}
	{\frac{m}{\gamma} (\kappa_\mathrm{C} +\kappa_\mathrm{H})^2/2 + \gamma(\kappa_\mathrm{C}+\kappa_\mathrm{H}+2\kappa)}  
	\geq 0,\label{eq:Stot2} 
\end{align}
which is also given in \eqref{eq:Stot} in the Main Text.
As expected, the total entropy production rate of the process is always greater or equal than zero. We further note that the rates of  entropy production and heat dissipated have contributions from the inertial terms (i.e., it depends on $m$). 
Figure~\ref{fig:heat2}b in the Main Text shows the  rate of entropy production from Eq.~\eqref{eq:Stot2} and from the MD simulations, with the latter given by the heat flows extracted by the thermostat divided by the respective temperatures.

\section{Detailed balance at the ``Pseudo Equilibrium'' point}
As described in the Main Text, we find that at $\kappa_\mathrm{C}T_\mathrm{H}=\kappa_\mathrm{H}T_\mathrm{C}$ the average values of the heat flows and the entropy production rate all vanish. 
To further investigate this ``pseudo equilibrium'' point, we check whether in this point 
the Langevin Eq.~(\ref{eq:LE-Matrix}) fulfills
detailed balance, meaning that all probability flows vanish. Detailed balance 
is a fundamental law  defining of systems in thermal equilibrium. We employ a reasoning inspired by the arguments used in \cite{Weiss2003}.

To this end, we consider the 
flow of the 
$4$-point joint probability density function (pdf), $\rho_{4}(\mathbf{x},\mathbf{v},t )$, of $\mathbf{x}=(x_\mathrm{C},x_\mathrm{H})^{{\mathsf{T}}}$
and $\mathbf{v}=(\dot x_\mathrm{C},\dot x_\mathrm{H})^{{\mathsf{T}}}$,
appearing in the corresponding {multivariate} Fokker-Planck equation. The Fokker-Planck equation reads
\begin{align}\label{eq:FPE-vector}
	\partial_t \rho_{4}(\mathbf{x},\mathbf{v},t) 
	=  
	-\nabla_x \underbrace{
		[\mathbf{v} \rho_{4}(\mathbf{x},\mathbf{v},t)]
	}_{=\mathbf{J}_x}, -\nabla_v \underbrace{
		[(\mathbf{A} \mathbf{x} - \mathbf{\Gamma}\mathbf{v})/m- \mathbf{D}_2\nabla_v ]\rho_{4}(\mathbf{x},\mathbf{v},t)
	}_{=\mathbf{J}_v},
\end{align} 
with the probability currents $\mathbf{J}_v,\mathbf{J}_x$ and the diffusion, friction and coupling matrices defined in \eqref{eq:LE-Matrix}. In general, the latter are constant in steady states, and zero in equilibrium. 
Now we use the identity $ \partial_z \rho= [\partial_z \ln(\rho)] \rho$, and define the 
four-dimensional phase space velocity $\mathbf{u}$ to rewrite the Fokker-Planck equation~(\ref{eq:FPE-vector}) as
\begin{align}\label{eq:phaseSpace}
	\partial_t \rho_{4} =   -\nabla
	\left[\mathbf{u} \rho_{4}  \right],~~~
	\text{with}~~~\mathbf{u} = ( \mathbf{v}, \mathbf{u}_v )^{\mathsf{T}},~~~
	\mathbf{u}_v =  \mathbf{A}/m\,\mathbf{x}-\mathbf{\Gamma}/m\,\mathbf{v}- \mathbf{D}_2\nabla_v \ln \rho_{4}(\mathbf{x},\mathbf{v}).
\end{align} 
The phase space velocity is connected to the probability current by $\mathbf{J}=\mathbf{u}\rho_{4}$. 
Detailed balance is fulfilled if all probability flows vanish. Hence, it is fulfilled if all components of $\mathbf{u}$ vanish. From (\ref{eq:phaseSpace}) we, in turn, find that the phase space velocity vanishes if
$    \mathbf{D}_2^{-1}(\mathbf{A}\mathbf{x}-\mathbf{\Gamma}\mathbf{v}) = m \nabla_v \ln \rho_{4}$. Thus, $\mathbf{D}_2^{-1}(\mathbf{A}\mathbf{x}-\mathbf{\Gamma}\mathbf{v})$ must be the gradient of a scalar function. The latter is true if and only if 
\begin{equation}
	\nabla \times [\mathbf{D}_2^{-1}(\mathbf{A}\mathbf{x}-\mathbf{\Gamma}\mathbf{v})] = 0.
\end{equation} 

Now, since in our model, 
$\mathbf{A}$, $\mathbf{\Gamma}$, and $\mathbf{D}_2$ are independent of $\mathbf{x}$ and $\mathbf{\Gamma}$ and $\mathbf{D}_2$ are diagonal, the last condition is only fulfilled if $\nabla \times  \mathbf{D}_2^{-1}\mathbf{A}\mathbf{x} = 0$.  Concretely, inserting the coupling and diffusion matrix defined  in \eqref{eq:LE-Matrix} and below, this can only be satisfied, if
\begin{align}\label{eq:Fulfill-DB}
	~\kappa_\mathrm{C}T_\mathrm{H}=\kappa_\mathrm{H}T_\mathrm{C}.
\end{align}
In summary, this reasoning has revealed that the detailed balance is fulfilled if and only if \eqref{eq:Fulfill-DB}, which coincides with the ``equilibrium condition'' found from the entropy production rate.
We stress that this condition is irrespective of the friction coefficients, the trap stiffness and the mass of the nanoparticles, and has the same form when starting with the overdamped limit of the Langevin equations \cite{Loos2020b}.

\newpage


\section{COP of the heat pump}

\begin{figure}[H]
	\centering
	\includegraphics[width=0.44\textwidth]{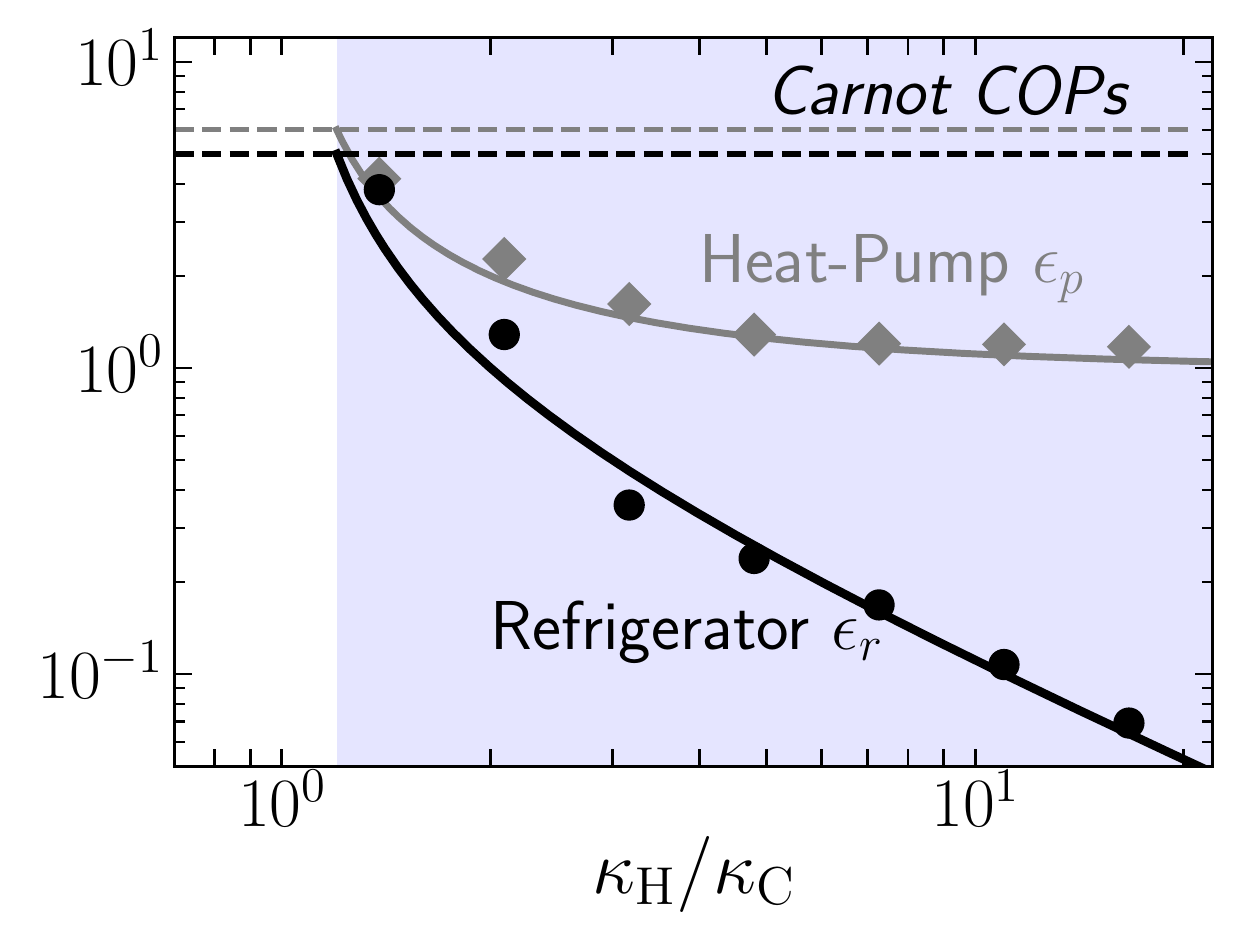}
	\caption{Coefficient of performance (COP) characterizing the mean efficiency of the refrigeration (black disks and line) and of the heat-pumping (grey diamonds and grey line) achieved by the nonreciprocally coupled nanoparticle system. Symbols stem from MD simulations, while the lines show the analytical predictions ($\langle \epsilon_p\rangle $ given in \eqref{eq:COP_pump}, $\epsilon_r$ given in the Main Text). Note the logarithmic axes. The horizontal dashed lines indicate the Carnot COPs, which in this case are 5 for the refrigerator and 6 for the heat pump. The Carnot values are reached in the limit $\kappa_\mathrm{H}\to \kappa_\mathrm{C}$. In this limit, the rate of extracted and pumped heat vanishes, however, vanishes, such that no heat is transferred in finite time.}\label{fig:COP-Ref+Pump}
\end{figure}

\begin{figure}[H]
	
	\begin{minipage}{0.29\textwidth}
		{\large{\textsf{(a)}}}\\
		\includegraphics[width=0.99\textwidth]{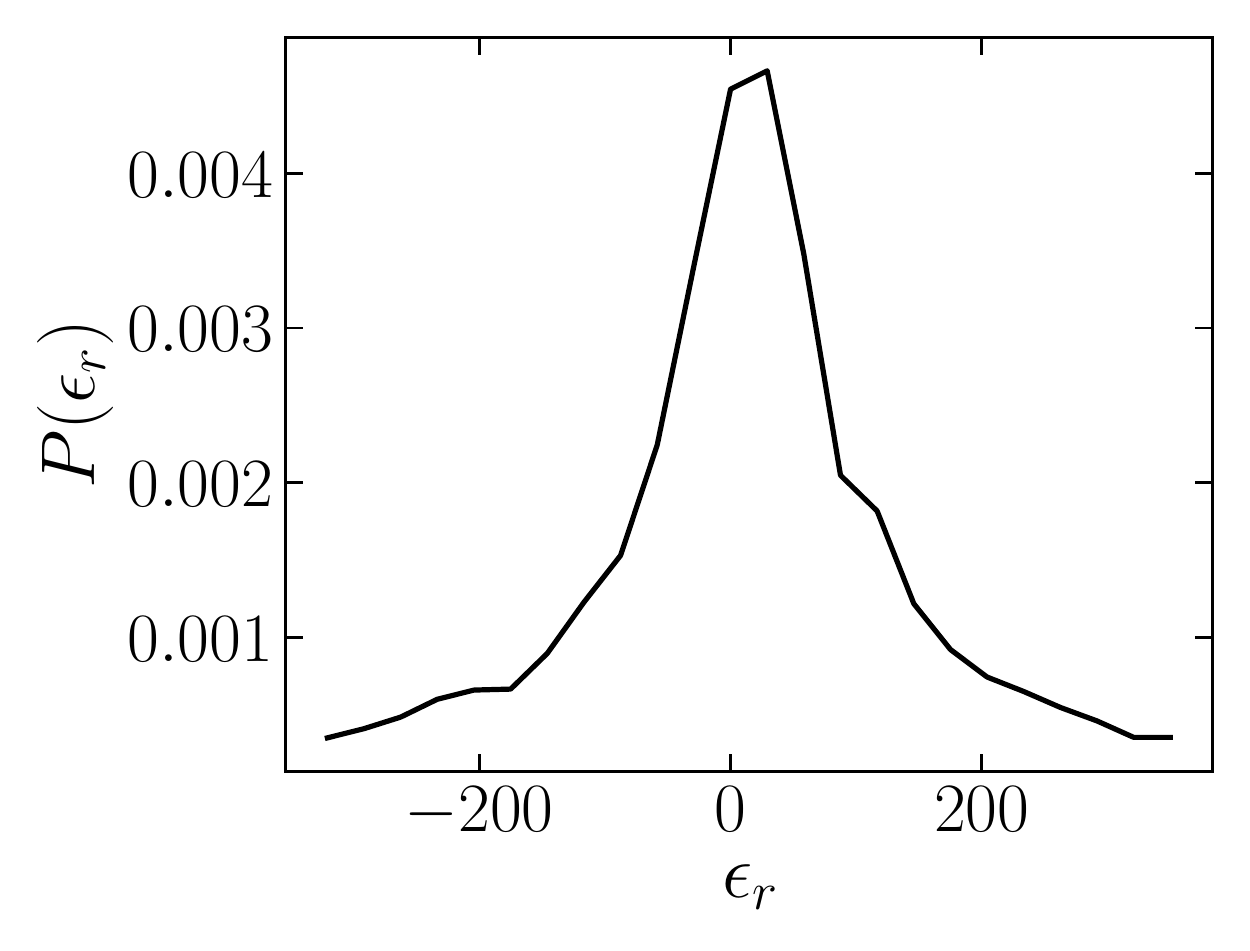}
	\end{minipage}
	\begin{minipage}{0.2\textwidth}
		{\large{\textsf{(b)}}}\\
		\includegraphics[width=0.89\textwidth]{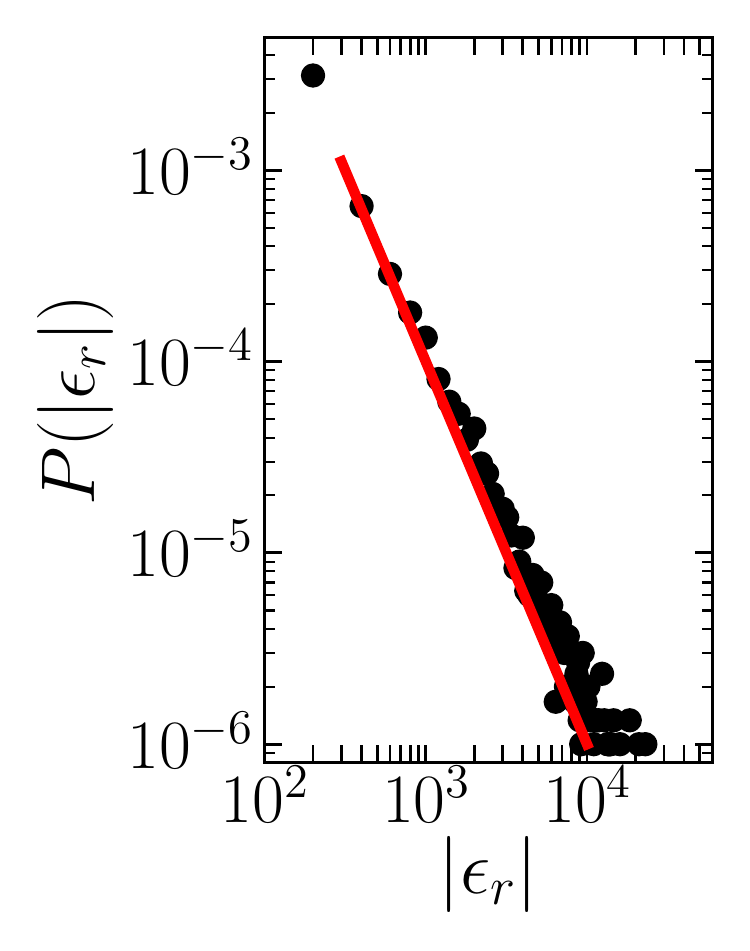}
	\end{minipage}	
	\begin{minipage}{0.29\textwidth}
		{\large{\textsf{(c)}}}\\
		\includegraphics[width=0.99\textwidth]{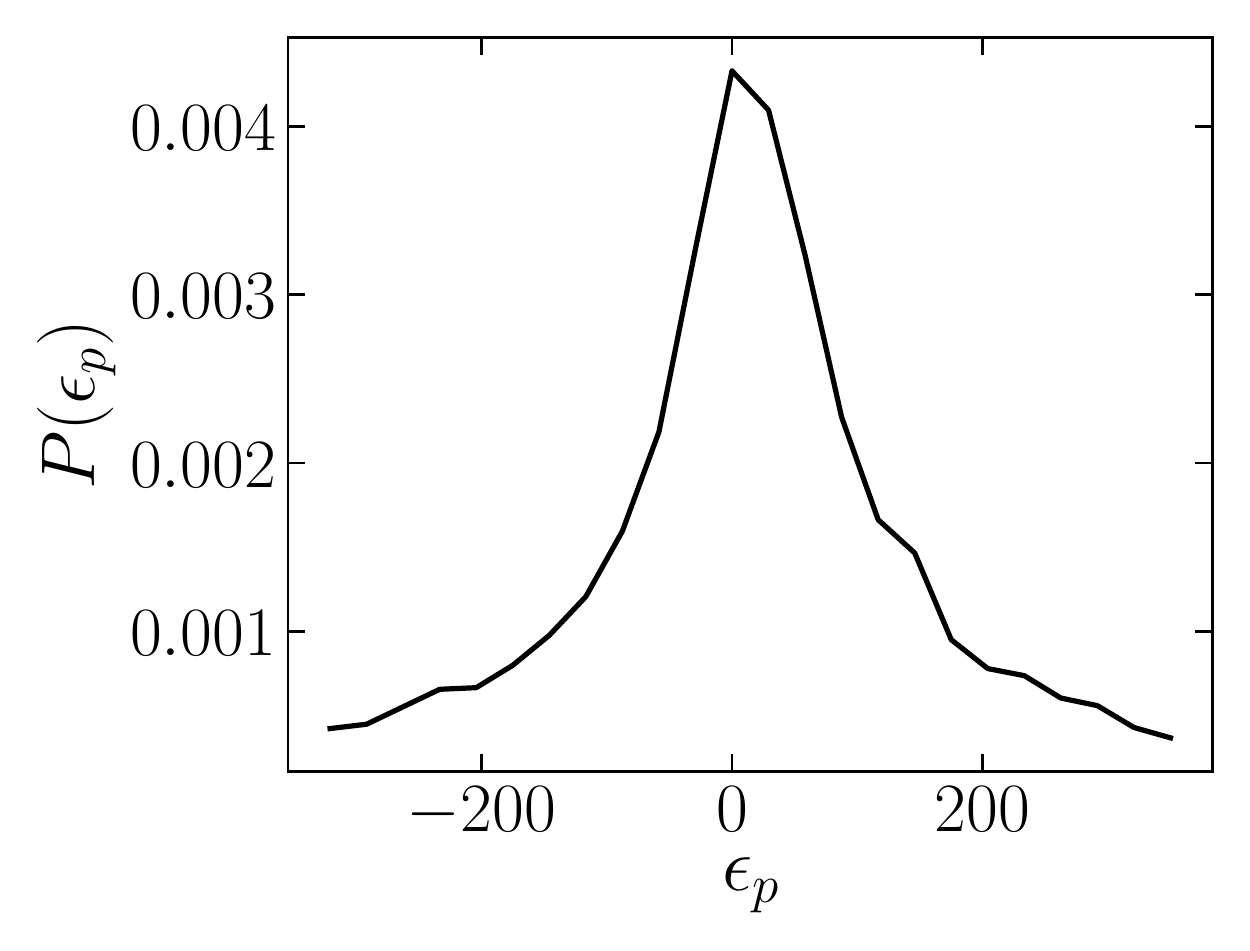}
	\end{minipage}	
	\begin{minipage}{0.2\textwidth}
		{\large{\textsf{(d)}}}\\
		\includegraphics[width=0.89\textwidth]{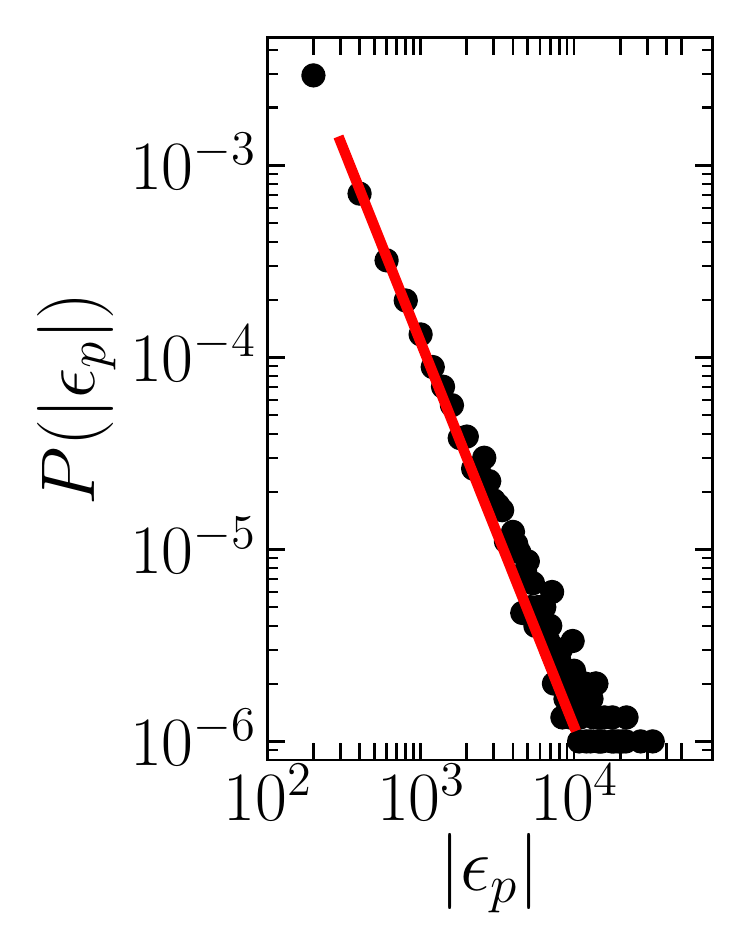}
	\end{minipage}

	\caption{
		(a,b) Fluctuations of the COP, $\epsilon_r = \frac{|\dot{Q}_\mathrm{C}|}{\dot{W}}$, for the refrigerator mode in linear (a) and logarithmic  scale (b). 
		(c,d) Fluctuations of the COP for the heat pump mode, 
		$\epsilon_p = \frac{|\dot{Q}_\mathrm{H}|}{\dot{W}}$,
		in linear (c) and logarithmic  scale (d). Note that the linear plots in (a,b) appears much more noisy, as few data points lie in the shown range, which is a consequence of the power-law distribution.
		The red lines in (c,d) depict power laws with exponent $-2$. The exponent of $-2$ is in good agreement with the COP of heat pump and of the refrigerator for all values of $\kappa_\mathrm{H}/\kappa_\mathrm{C}$ considered.  }\label{fig:COP-fluctuations}
\end{figure}

In the Main Text, we consider the system as a nano refrigerator, i.e., we quantify the efficiency of cooling. However, one can also use this machine as a heat pump to heat up the hot bath. The efficiency of heat pumping can be measured by an analogous coefficient of performance defined as
\begin{align}
	\langle \epsilon_p \rangle 
	= \frac{|\langle \dot{Q}_\mathrm{H}\rangle |}{\langle \dot{W}\rangle  }
	=
	\frac{|\langle\dot{Q}_\mathrm{H}\rangle|}{\langle\dot{Q}_\mathrm{H}\rangle+\langle\dot{Q}_\mathrm{C}\rangle }
	=
	\frac{\kappa_\mathrm{H}}
	{\kappa_\mathrm{H}-\kappa_\mathrm{C}}  .
	\label{eq:COP_pump}
\end{align}
Like the COP of the refrigerator mode, 
the COP of the heat pump is bounded by the Carnot value  $\langle \epsilon_p \rangle \leq \frac{T_\mathrm{H}}
{T_\mathrm{H}-T_\mathrm{C}} $. Figure \ref{fig:COP-Ref+Pump} displays both COPs for the refrigerator and heat pump, from simulations and from the theoretical predictions. Figure \ref{fig:COP-fluctuations} shows the distributions of the fluctuating COPs, which are discussed in the Main Text. 


\section{Derivation of the distribution of power [Eq.~\eqref{eq:Pw} in the Main Text]}
Here, we derive the distribution of the instantaneous (stochastic) power. We make use of the explicit expression for the stationary joint probability density for the position and velocities of the two nanoparticles $\rho_4(x_\mathrm{C},x_\mathrm{H},v_\mathrm{C},v_\mathrm{H})$, which is a four-variate Gaussian distribution with zero mean and covariance matrix $\mathbf{C}$ as given in Appendix~\ref{sec:heatCalculation}.

To access the distribution of $\dot W_j$ with $j \in \{\mathrm{C,H}\}$, we first recall that 
\begin{equation}
	\dot W_j =(\mathrm{d}t)^{-1} (\kappa_j x_l \circ \mathrm{d}x_j ) =  \kappa_j x_l \circ v_j,
\end{equation}
with $l \neq j$ which we assume throughout this section. From $\rho_4$, we can  derive $P(\dot W_j)$ by the following integral
\begin{align}
	P(\dot W_j) &= \int_{-\infty}^{\infty} \mathrm{d}x_j
	\int_{-\infty}^{\infty} \mathrm{d}x_l
	\int_{-\infty}^{\infty} \mathrm{d}v_j
	\int_{-\infty}^{\infty} \mathrm{d}v_l
	\,
	\rho_4(x_j,x_l,v_j,v_l) \delta (\dot{W}_j - \kappa_j x_l v_j) 
	\\
	&=
	\int_{-\infty}^{\infty} \mathrm{d}x_l
	\int_{-\infty}^{\infty} \mathrm{d}v_j
	\,
	\rho_2(x_l,v_j) \delta (\dot{W}_j - \kappa_j x_l v_j).\label{eq:e3}
\end{align}
Here,
\begin{equation}\label{eq:rho2}
	\rho_2(x_l,v_j)  = \int_{-\infty}^{\infty} \text{d}x_j \int_{-\infty}^{\infty}\text{d}v_l \rho_4(x_j,x_l,v_j,v_l)
	=
	\mathcal{N}_j^{-1} \exp(-\alpha_j x_l^2 + \beta_j x_l v_j - \zeta_j v_j^2),
\end{equation}
is the joint stationary distribution of the position of the $l-$nanoparticle and the velocity of the $j-$nanoparticle, with $j\neq l$. It is given by a bivariate normal distribution 
with normalization constant 
${\mathcal{N}_j}$,
and coefficients 
\begin{align}\label{def:2pointpdf}
	\alpha_j &= \frac{1}{2 \langle X_l^2 \rangle (1-\psi_j^2)},
	~~
	\beta_j = \frac{\psi_j}{\sqrt{\langle  X_l^2 \rangle\langle \dot X_j^2 \rangle}(1-\psi^2_j)},
	~~\zeta_j = \frac{1}{2 \langle \dot X_j^2 \rangle (1-\psi_j^2)}, ~~
	\psi_j=\frac{\langle X_l \dot{X}_j \rangle }{\sqrt{\langle  X_l^2 \rangle\langle \dot X_j^2 \rangle}},
\end{align}
which are functions of the covariances of the position and velocities of the nanoparticles. 
Therefore, the distribution of the power depends on the functions $\alpha_j,\beta_j,\zeta_j$ defined in \eqref{def:2pointpdf}, which in turn depend on $\langle X_l^2 \rangle$, $\langle X_l\dot{X}_j \rangle$ and $\langle\dot{X}_j^2 \rangle$.  See Appendix~\ref{sec:heatflowrates} for explicit analytical expressions of the covariances in terms of the physical parameters of the system.

Using the explicit expression of $\rho_2$ given by Eq.~\eqref{eq:rho2} and integrating out the delta distribution in Eq.~\eqref{eq:e3}, we obtain
\begin{align}
	P(\dot W_j) &= 
	\int_{-\infty}^{\infty} \mathrm{d}x_l
	\int_{-\infty}^{\infty} \mathrm{d}v_j
	\rho_2(x_l,v_j) ~\delta \left(\frac{\dot{W}_j}{\kappa_j x_l} - v_j\right)
	=
	\int_{-\infty}^{\infty} \mathrm{d}x_l
	\rho_2\left(x_l,\frac{\dot{W}_j}{\kappa_j x_l}\right) 
	\\
	&=
	\frac{1}{\mathcal{N}_j} \,\int_{-\infty}^{\infty} \mathrm{d}x_l
	\exp\left(-\alpha_j x_l^2 +\beta_j x_l \frac{\dot{W}_j}{\kappa_j x_l} - \zeta_j \frac{\dot{W}_j^2}{\kappa_j^2 x_l^2}\right)
	=
	\frac{1}{\mathcal{N}_j} \,\exp\left( \frac{\beta_j}{\kappa_j}  \dot{W}_j \right)\int_{-\infty}^{\infty} \mathrm{d}x_l
	\exp\left(-\alpha_j x_l^2 -  \frac{\zeta_j \dot{W}_j^2}{\kappa_j^2} \frac{1}{x_l^2}\right).
\end{align}
Changing variables  $y=\sqrt{\alpha_j}x_l$  and absorbing  the Jacobian of the transformation in the new normalization constant $\mathcal{Z}_j=\alpha_j^2\mathcal{N}_j$   yields
\begin{align}
	P(\dot W_j) = \frac{1}{\mathcal{N}_j} \exp\left( \frac{\beta_j}{\kappa_j}  \dot{W}_j \right) \int_{-\infty}^{\infty} \frac{\mathrm{d}y}{\alpha_j^2}\,
	\exp\left(-y^2 -  \frac{\zeta_j\dot{W}_j^2}{\kappa_j^2} \frac{\alpha_j}{y^2}\right) 
	=
	\frac{1}{\mathcal{Z}_j} \exp\left(  \frac{\beta_j}{\kappa_j}  \dot{W}_j \right) \int_{-\infty}^{\infty} \mathrm{d}y\,
	\exp\left(-y^2 -  \frac{b}{y^2}\right) .\label{eq:e8}
\end{align}
with $b\equiv  (\zeta_j \alpha_j\dot{W}_j^2)/\kappa_j^2$.
Applying  the property $\int_{-\infty}^{\infty} \mathrm{d}y\,
\exp\left(-y^2 -  \frac{b}{y^2}\right) =
\sqrt{\pi}\exp(-2\sqrt{b})$ which holds  for all $d>0$, Eq.~\eqref{eq:e8}  yields the distribution of the power exerted on the $j \in \{\mathrm{C,H}\}$ nanoparticle
\begin{equation}\label{eq:Pwappendix} 
	P(\dot W_j)= 
	\frac{1}{\mathcal{Z}_j}
	\exp
	\left[ 
	\frac{\beta_j}{\kappa_j}  \dot{W}_j -2\frac{\sqrt{\zeta \alpha_j}}{|\kappa_j|} |\dot{W}_j| \right],
\end{equation}
which coincides with Eq.~\eqref{eq:Pw} in the Main Text.

\begin{figure}[H]
	{\large{\textsf{(a)}}}~~~~~~~~~~~~~~~~~~~~~~~~~~~~~~~~~~~~~~~~~~~~~~~~~~~~~~~~~~~~~~~~~~~~~~~~~~~~~~~~~~~~~~~~~~~~~~~~~~~{\large{\textsf{(b)}}}\\
	\includegraphics[width=0.4\textwidth]{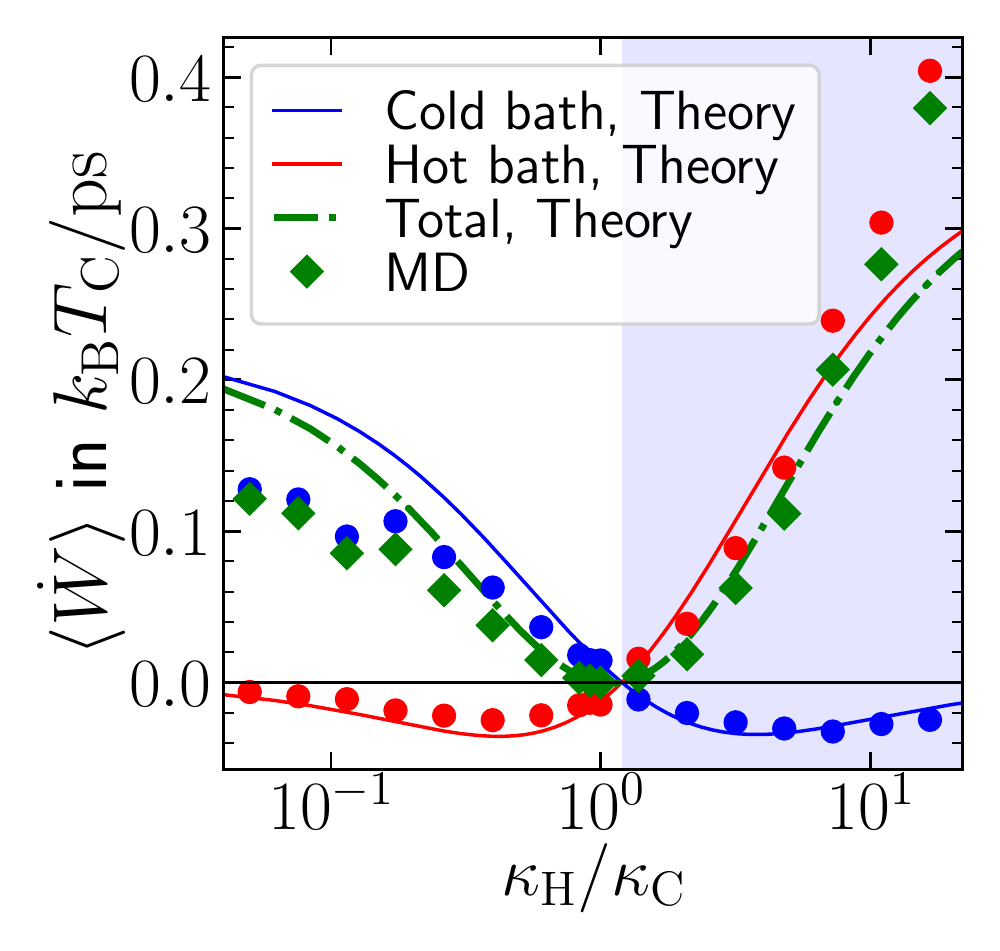}~~~~~~~~~~~~~~~~~~~~
	\includegraphics[width=0.4\textwidth]{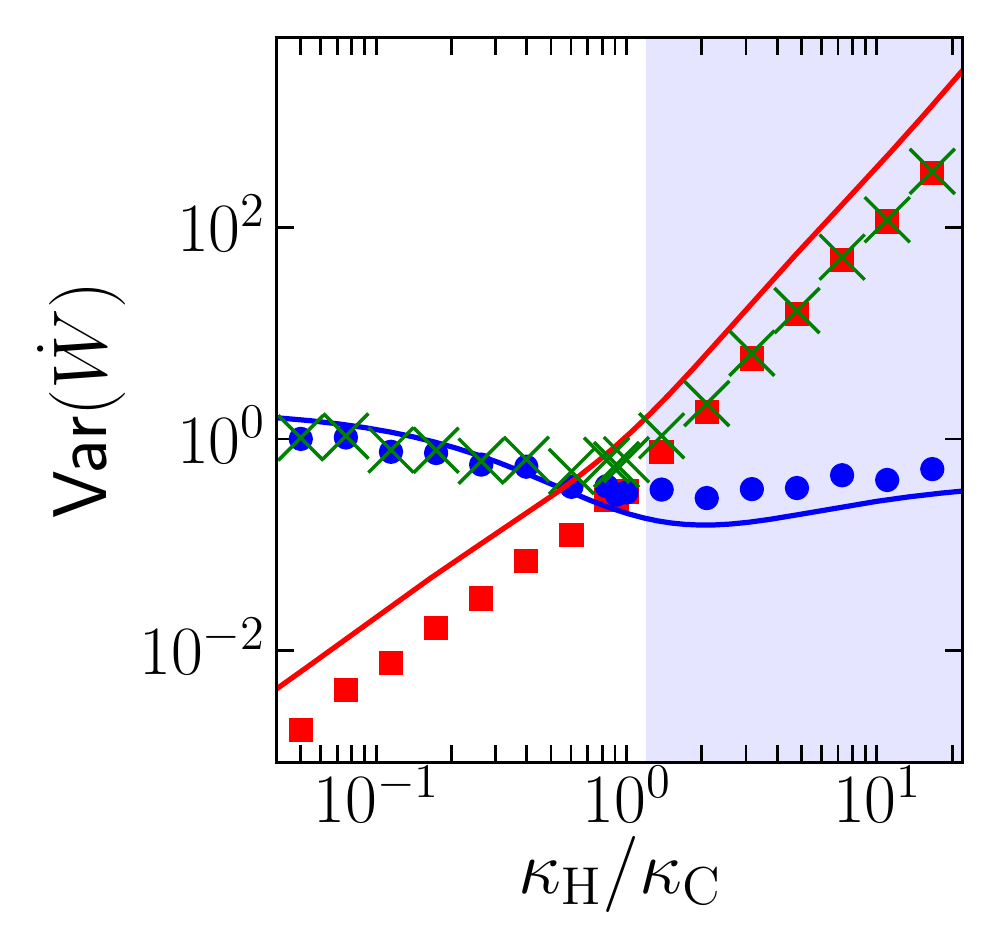}
	\caption{Mean value (a) and variance (b) of the  power exerted on the nanoparticles obtained from MD simulations (symbols) and according to the theoretical prediction (lines), for the cold particle ($\dot{W}_\mathrm{C}$, blue), the hot particle ($\dot{W}_\mathrm{H}$, red) and the total power ($\dot{W}=\dot{W}_\mathrm{C}+\dot{W}_\mathrm{H}$, green). }\label{fig:PowerFluctuations}
\end{figure}

From the  distributions~\eqref{eq:Pwappendix}, we can further deduce analytical expressions for any moment of the power. In particular, the variance of the power reads:
\begin{align}
	\text{Var}(\dot W_j)& =\left( \int_{-\infty}^{\infty} P(\dot W_j) \dot W_j\right)^2 -  \int_{-\infty}^{\infty} P(\dot W_j) \dot W_j^2\nonumber\\
	& = 
	\frac{1}{\mathcal{Z}}\!\left[ \frac{2\kappa_j^3}{(2\sqrt{\alpha_j \zeta_j}+\beta_j)^3} +
	\frac{2\kappa_j^3}{(2\sqrt{\alpha_j \zeta_j}-\beta_j)^3}
	-
	\left( \frac{\kappa_j^2}{(2\sqrt{\alpha_j \zeta_j}+\beta_j)^2} +
	\frac{2\kappa_j^2}{(2\sqrt{\alpha_j \zeta_j}-\beta_j)^2}\right)^2 \right] .\label{eq:varworktheory}
\end{align}
Figure \ref{fig:PowerFluctuations} depicts the mean and the variance of the power fluctuations. We find good qualitative agreement between theory [Eq.~(\ref{eq:varworktheory})] and the MD simulation results.

\section{Stability conditions}
Here we study the stability of the process described by the Langevin equation~\eqref{eq:Le}, using the matrix equation given in \eqref{eq:LE-Matrix}.
The stability conditions can be found from the roots of the eigenvalues of the generalised coupling matrix $-\mathbf{B}$. The eigenvalues read 
\begin{align}
	\lambda_{1/2}=&\frac{1}{2m} \left(-\gamma \pm \sqrt{\gamma^2-4m\kappa}\right),\\
	\lambda_{3/4}=&\frac{1}{2m} \left(-\gamma \pm \sqrt{\gamma^2-4m(\kappa+\kappa_\mathrm{H} + \kappa_\mathrm{C})}\right).
\end{align}
The two largest eigenvalues are those with the respective plus sign. From them, we find the stability conditions $-\gamma + \sqrt{\gamma^2-4m \kappa}<0$, thus $  \kappa > 0 $, and $-\gamma + \sqrt{\gamma^2-4m(\kappa+\kappa_\mathrm{H} + \kappa_\mathrm{C})}<0$, 
thus
$ \kappa_\mathrm{H} + \kappa_\mathrm{C} > -\kappa. $ To summarize, the system is stable if
\begin{align}
	\kappa > 0,~\text{and}~~  \kappa_\mathrm{H} + \kappa_\mathrm{C} > -\kappa .
\end{align}
We note that these conditions are independent of $m$ and coincide with the stability boundaries of the overdamped system (as expected). Further note that the eigenvalues become complex if $\gamma^2 < 4m \kappa$, or $\gamma^2 < 4m(\kappa+\kappa_\mathrm{C}+\kappa_\mathrm{H})$. For the parameter choices considered in the MD simulations, all eigenvalues have imaginary parts, indicating oscillatory behaviour (for all considered $\kappa_\mathrm{H}, \kappa_\mathrm{C}$ values). The oscillatory character of the stochastic dynamics (which stems from the inertial terms) manifests itself in the negative values in the velocity autocorrelation function (see Fig.~\ref{fig:PACF}).
%

%

\end{document}